
\input psfig.sty
%
%
%
%

\catcode `\@=11 

\def\@version{1.4}
\def\@verdate{22nd Feb 1994}

%
%
%
%


\newif\ifprod@font

\ifx\@typeface\undefined
  \def\@typeface{Comp. Modern}\prod@fontfalse
\else
  \prod@fonttrue 
\fi

\def\newfam{\alloc@8\fam\chardef\sixt@@n} 

\ifprod@font
\font\fiverm=mtr10 at 5pt
\font\fivebf=mtbx10 at 5pt
\font\fiveit=mtti10 at 5pt
\font\fivesl=mtsl10 at 5pt
\font\fivett=mttt10 at 5pt     \hyphenchar\fivett=-1
\font\fivecsc=mtcsc10 at 5pt
\font\fivesf=mtss10 at 5pt
\font\fivei=mtmi10 at 5pt      \skewchar\fivei='177
\font\fivemib=mtmib10 at 5pt   \skewchar\fivemib='177
\font\fivesy=mtsy10 at 5pt     \skewchar\fivesy='60
\font\fivesyb=mtbsy10 at 5pt   \skewchar\fivesyb='60

\font\sixrm=mtr10 at 6pt
\font\sixbf=mtbx10 at 6pt
\font\sixit=mtti10 at 6pt
\font\sixsl=mtsl10 at 6pt
\font\sixtt=mttt10 at 6pt      \hyphenchar\sixtt=-1
\font\sixcsc=mtcsc10 at 6pt
\font\sixsf=mtss10 at 6pt
\font\sixi=mtmi10 at 6pt       \skewchar\sixi='177
\font\sixmib=mtmib10 at 6pt    \skewchar\sixmib='177
\font\sixsy=mtsy10 at 6pt      \skewchar\sixsy='60
\font\sixsyb=mtbsy10 at 6pt    \skewchar\sixsyb='60

\font\sevenrm=mtr10 at 7pt
\font\sevenbf=mtbx10 at 7pt
\font\sevenit=mtti10 at 7pt
\font\sevensl=mtsl10 at 7pt
\font\seventt=mttt10 at 7pt     \hyphenchar\seventt=-1
\font\sevencsc=mtcsc10 at 7pt
\font\sevensf=mtss10 at 7pt
\font\seveni=mtmi10 at 7pt      \skewchar\seveni='177
\font\sevenmib=mtmib10 at 7pt   \skewchar\sevenmib='177
\font\sevensy=mtsy10 at 7pt     \skewchar\sevensy='60
\font\sevensyb=mtbsy10 at 7pt   \skewchar\sevensyb='60

\font\eightrm=mtr10 at 8pt
\font\eightbf=mtbx10 at 8pt
\font\eightit=mtti10 at 8pt
\font\eighti=mtmi10 at 8pt      \skewchar\eighti='177
\font\eightmib=mtmib10 at 8pt   \skewchar\eightmib='177
\font\eightsy=mtsy10 at 8pt     \skewchar\eightsy='60
\font\eightsyb=mtbsy10 at 8pt   \skewchar\eightsyb='60
\font\eightsl=mtsl10 at 8pt
\font\eighttt=mttt10 at 8pt     \hyphenchar\eighttt=-1
\font\eightcsc=mtcsc10 at 8pt
\font\eightsf=mtss10 at 8pt

\font\ninerm=mtr10 at 9pt
\font\ninebf=mtbx10 at 9pt
\font\nineit=mtti10 at 9pt
\font\ninei=mtmi10 at 9pt      \skewchar\ninei='177
\font\ninemib=mtmib10 at 9pt   \skewchar\ninemib='177
\font\ninesy=mtsy10 at 9pt     \skewchar\ninesy='60
\font\ninesyb=mtbsy10 at 9pt   \skewchar\ninesyb='60
\font\ninesl=mtsl10 at 9pt
\font\ninett=mttt10 at 9pt     \hyphenchar\ninett=-1
\font\ninecsc=mtcsc10 at 9pt
\font\ninesf=mtss10 at 9pt

\font\tenrm=mtr10
\font\tenbf=mtbx10
\font\tenit=mtti10
\font\teni=mtmi10		\skewchar\teni='177
\font\tenmib=mtmib10	\skewchar\tenmib='177
\font\tensy=mtsy10		\skewchar\tensy='60
\font\tensyb=mtbsy10	\skewchar\tensyb='60
\font\tenex=cmex10
\font\tensl=mtsl10
\font\tentt=mttt10		\hyphenchar\tentt=-1
\font\tencsc=mtcsc10
\font\tensf=mtss10

\font\elevenrm=mtr10 at 11pt
\font\elevenbf=mtbx10 at 11pt
\font\elevenit=mtti10 at 11pt
\font\eleveni=mtmi10 at 11pt      \skewchar\eleveni='177
\font\elevenmib=mtmib10 at 11pt   \skewchar\elevenmib='177
\font\elevensy=mtsy10 at 11pt     \skewchar\elevensy='60
\font\elevensyb=mtbsy10 at 11pt   \skewchar\elevensyb='60
\font\elevensl=mtsl10 at 11pt
\font\eleventt=mttt10 at 11pt     \hyphenchar\eleventt=-1
\font\elevencsc=mtcsc10 at 11pt
\font\elevensf=mtss10 at 11pt

\font\twelverm=mtr10 at 12pt
\font\twelvebf=mtbx10 at 12pt
\font\twelveit=mtti10 at 12pt
\font\twelvesl=mtsl10 at 12pt
\font\twelvett=mttt10 at 12pt     \hyphenchar\twelvett=-1
\font\twelvecsc=mtcsc10 at 12pt
\font\twelvesf=mtss10 at 12pt
\font\twelvei=mtmi10 at 12pt      \skewchar\twelvei='177
\font\twelvemib=mtmib10 at 12pt   \skewchar\twelvemib='177
\font\twelvesy=mtsy10 at 12pt     \skewchar\twelvesy='60
\font\twelvesyb=mtbsy10 at 12pt   \skewchar\twelvesyb='60

\font\fourteenrm=mtr10 at 14pt
\font\fourteenbf=mtbx10 at 14pt
\font\fourteenit=mtti10 at 14pt
\font\fourteeni=mtmi10 at 14pt      \skewchar\fourteeni='177
\font\fourteenmib=mtmib10 at 14pt   \skewchar\fourteenmib='177
\font\fourteensy=mtsy10 at 14pt     \skewchar\fourteensy='60
\font\fourteensyb=mtbsy10 at 14pt   \skewchar\fourteensyb='60
\font\fourteensl=mtsl10 at 14pt
\font\fourteentt=mttt10 at 14pt     \hyphenchar\fourteentt=-1
\font\fourteencsc=mtcsc10 at 14pt
\font\fourteensf=mtss10 at 14pt

\font\seventeenrm=mtr10 at 17pt
\font\seventeenbf=mtbx10 at 17pt
\font\seventeenit=mtti10 at 17pt
\font\seventeeni=mtmi10 at 17pt      \skewchar\seventeeni='177
\font\seventeenmib=mtmib10 at 17pt   \skewchar\seventeenmib='177
\font\seventeensy=mtsy10 at 17pt     \skewchar\seventeensy='60
\font\seventeensyb=mtbsy10 at 17pt   \skewchar\seventeensyb='60
\font\seventeensl=mtsl10 at 17pt
\font\seventeentt=mttt10 at 17pt     \hyphenchar\seventeentt=-1
\font\seventeencsc=mtcsc10 at 17pt
\font\seventeensf=mtss10 at 17pt


\newfam\xmfam
\newfam\ymfam

\font\fivexm=mtxm10 at 5pt
\font\sixxm=mtxm10 at 6pt
\font\sevenxm=mtxm10 at 7pt
\font\eightxm=mtxm10 at 8pt
\font\ninexm=mtxm10 at 9pt
\font\tenxm=mtxm10
\font\elevenxm=mtxm10 at 11pt
\font\twelvexm=mtxm10 at 12pt
\font\fourteenxm=mtxm10 at 14pt
\font\seventeenxm=mtxm10 at 17pt

\font\fiveym=mtym10 at 5pt
\font\sixym=mtym10 at 6pt
\font\sevenym=mtym10 at 7pt
\font\eightym=mtym10 at 8pt
\font\nineym=mtym10 at 9pt
\font\tenym=mtym10
\font\elevenym=mtym10 at 11pt
\font\twelveym=mtym10 at 12pt
\font\fourteenym=mtym10 at 14pt
\font\seventeenym=mtym10 at 17pt
\else
\font\fiverm=cmr5
\font\fivei=cmmi5             \skewchar\fivei='177
\font\fivemib=cmmib10 at 5pt  \skewchar\fivemib='177
\font\fivesy=cmsy5            \skewchar\fivesy='60
\font\fivesyb=cmbsy10 at 5pt  \skewchar\fivesyb='60
\font\fivebf=cmbx5

\font\sixrm=cmr6
\font\sixi=cmmi6             \skewchar\sixi='177
\font\sixmib=cmmib10 at 6pt  \skewchar\sixmib='177
\font\sixsy=cmsy6            \skewchar\sixsy='60
\font\sixsyb=cmbsy10 at 6pt  \skewchar\sixsyb='60
\font\sixbf=cmbx6

\font\sevenrm=cmr7
\font\seveni=cmmi7             \skewchar\seveni='177
\font\sevenmib=cmmib10 at 7pt  \skewchar\sevenmib='177
\font\sevensy=cmsy7            \skewchar\sevensy='60
\font\sevensyb=cmbsy10 at 7pt  \skewchar\sevensyb='60
\font\sevenbf=cmbx7

\font\eightrm=cmr8
\font\eightbf=cmbx8
\font\eightit=cmti8
\font\eighti=cmmi8			\skewchar\eighti='177
\font\eightmib=cmmib10 at 8pt	\skewchar\eightmib='177
\font\eightsy=cmsy8			\skewchar\eightsy='60
\font\eightsyb=cmbsy10 at 8pt	\skewchar\eightsyb='60
\font\eightsl=cmsl8
\font\eighttt=cmtt8			\hyphenchar\eighttt=-1
\font\eightcsc=cmcsc10 at 8pt
\font\eightsf=cmss8

\font\ninerm=cmr9
\font\ninebf=cmbx9
\font\nineit=cmti9
\font\ninei=cmmi9			\skewchar\ninei='177
\font\ninemib=cmmib10 at 9pt	\skewchar\ninemib='177
\font\ninesy=cmsy9			\skewchar\ninesy='60
\font\ninesyb=cmbsy10 at 9pt	\skewchar\ninesyb='60
\font\ninesl=cmsl9
\font\ninett=cmtt9			\hyphenchar\ninett=-1
\font\ninecsc=cmcsc10 at 9pt
\font\ninesf=cmss9

\font\tenrm=cmr10
\font\tenbf=cmbx10
\font\tenit=cmti10
\font\teni=cmmi10		\skewchar\teni='177
\font\tenmib=cmmib10	\skewchar\tenmib='177
\font\tensy=cmsy10		\skewchar\tensy='60
\font\tensyb=cmbsy10	\skewchar\tensyb='60
\font\tenex=cmex10
\font\tensl=cmsl10
\font\tentt=cmtt10		\hyphenchar\tentt=-1
\font\tencsc=cmcsc10
\font\tensf=cmss10

\font\elevenrm=cmr10 scaled \magstephalf
\font\elevenbf=cmbx10 scaled \magstephalf
\font\elevenit=cmti10 scaled \magstephalf
\font\eleveni=cmmi10 scaled \magstephalf	\skewchar\eleveni='177
\font\elevenmib=cmmib10 scaled \magstephalf	\skewchar\elevenmib='177
\font\elevensy=cmsy10 scaled \magstephalf	\skewchar\elevensy='60
\font\elevensyb=cmbsy10 scaled \magstephalf	\skewchar\elevensyb='60
\font\elevensl=cmsl10 scaled \magstephalf
\font\eleventt=cmtt10 scaled \magstephalf	\hyphenchar\eleventt=-1
\font\elevencsc=cmcsc10 scaled \magstephalf
\font\elevensf=cmss10 scaled \magstephalf

\font\twelverm=cmr10 scaled \magstep1
\font\twelvebf=cmbx10 scaled \magstep1
\font\twelvei=cmmi10 scaled \magstep1      \skewchar\twelvei='177
\font\twelvemib=cmmib10 scaled \magstep1   \skewchar\twelvemib='177
\font\twelvesy=cmsy10 scaled \magstep1     \skewchar\twelvesy='60
\font\twelvesyb=cmbsy10 scaled \magstep1   \skewchar\twelvesyb='60

\font\fourteenrm=cmr10 scaled \magstep2
\font\fourteenbf=cmbx10 scaled \magstep2
\font\fourteenit=cmti10 scaled \magstep2
\font\fourteeni=cmmi10 scaled \magstep2		\skewchar\fourteeni='177
\font\fourteenmib=cmmib10 scaled \magstep2	\skewchar\fourteenmib='177
\font\fourteensy=cmsy10 scaled \magstep2	\skewchar\fourteensy='60
\font\fourteensyb=cmbsy10 scaled \magstep2	\skewchar\fourteensyb='60
\font\fourteensl=cmsl10 scaled \magstep2
\font\fourteentt=cmtt10 scaled \magstep2	\hyphenchar\fourteentt=-1
\font\fourteencsc=cmcsc10 scaled \magstep2
\font\fourteensf=cmss10 scaled \magstep2

\font\seventeenrm=cmr10 scaled \magstep3
\font\seventeenbf=cmbx10 scaled \magstep3
\font\seventeenit=cmti10 scaled \magstep3
\font\seventeeni=cmmi10 scaled \magstep3	\skewchar\seventeeni='177
\font\seventeenmib=cmmib10 scaled \magstep3	\skewchar\seventeenmib='177
\font\seventeensy=cmsy10 scaled \magstep3	\skewchar\seventeensy='60
\font\seventeensyb=cmbsy10 scaled \magstep3	\skewchar\seventeensyb='60
\font\seventeensl=cmsl10 scaled \magstep3
\font\seventeentt=cmtt10 scaled \magstep3	\hyphenchar\seventeentt=-1
\font\seventeencsc=cmcsc10 scaled \magstep3
\font\seventeensf=cmss10 scaled \magstep3
\fi

\def\hexnumber#1{\ifcase#1 0\or1\or2\or3\or4\or5\or6\or7\or8\or9\or
  A\or B\or C\or D\or E\or F\fi}

\ifprod@font
  \edef\@xm{\hexnumber\xmfam}
  \edef\@ym{\hexnumber\ymfam}
\fi

\def\makestrut{%
  \setbox\strutbox=\hbox{%
    \vrule height.7\baselineskip depth.3\baselineskip width \z@}%
}

\def\baselinestretch{1}
\newskip\tmp@bls

\def\b@ls#1{
  \tmp@bls=#1\relax
  \baselineskip=#1\relax\makestrut
  \normalbaselineskip=\baselinestretch\tmp@bls
  \normalbaselines
}

\def\nostb@ls#1{
  \normalbaselineskip=#1\relax
  \normalbaselines
  \makestrut
}

%

\newfam\mibfam 
\newfam\sybfam 
\newfam\scfam  
\newfam\sffam  

\def\mit{\fam\@ne}

\def\cal{\fam\tw@}

\def\em{\ifdim\fontdimen1\font>\z@ \rm\else\it\fi}

\textfont3=\tenex
\scriptfont3=\tenex
\scriptscriptfont3=\tenex

\setbox0=\hbox{\tenex B} \p@renwd=\wd0 

\def\eightpoint{
  \def\rm{\fam0\eightrm}%
  \textfont0=\eightrm \scriptfont0=\sixrm \scriptscriptfont0=\fiverm%
  \textfont1=\eighti  \scriptfont1=\sixi  \scriptscriptfont1=\fivei%
  \textfont2=\eightsy \scriptfont2=\sixsy \scriptscriptfont2=\fivesy%
  \textfont\itfam=\eightit\def\it{\fam\itfam\eightit}%
  \ifprod@font
    \scriptfont\itfam=\sixit
      \scriptscriptfont\itfam=\fiveit
  \else
    \scriptfont\itfam=\eightit
      \scriptscriptfont\itfam=\eightit
  \fi
  \textfont\bffam=\eightbf%
    \scriptfont\bffam=\sixbf%
      \scriptscriptfont\bffam=\fivebf%
  \def\bf{\fam\bffam\eightbf}%
  \textfont\slfam=\eightsl\def\sl{\fam\slfam\eightsl}%
  \ifprod@font
    \scriptfont\slfam=\sixsl
      \scriptscriptfont\slfam=\fivesl
  \else
    \scriptfont\slfam=\eightsl
      \scriptscriptfont\slfam=\eightsl
  \fi
  \textfont\ttfam=\eighttt\def\tt{\fam\ttfam\eighttt}%
  \ifprod@font
    \scriptfont\ttfam=\sixtt
      \scriptscriptfont\ttfam=\fivett
  \else
    \scriptfont\ttfam=\eighttt
      \scriptscriptfont\ttfam=\eighttt
  \fi
  \textfont\scfam=\eightcsc\def\sc{\fam\scfam\eightcsc}%
  \ifprod@font
    \scriptfont\scfam=\sixcsc
      \scriptscriptfont\scfam=\fivecsc
  \else
    \scriptfont\scfam=\eightcsc
      \scriptscriptfont\scfam=\eightcsc
  \fi
  \textfont\sffam=\eightsf\def\sf{\fam\sffam\eightsf}%
  \ifprod@font
    \scriptfont\sffam=\sixsf
      \scriptscriptfont\sffam=\fivesf
  \else
    \scriptfont\sffam=\eightsf
      \scriptscriptfont\sffam=\eightsf
  \fi
  \textfont\mibfam=\eightmib
    \scriptfont\mibfam=\sixmib
      \scriptscriptfont\mibfam=\fivemib
  \textfont\sybfam=\eightsyb
    \scriptfont\sybfam=\sixsyb
      \scriptscriptfont\sybfam=\fivesyb
  \ifprod@font
    \textfont\xmfam=\eightxm
      \scriptfont\xmfam=\sixxm
        \scriptscriptfont\xmfam=\fivexm
    \textfont\ymfam=\eightym
      \scriptfont\ymfam=\sixym
        \scriptscriptfont\ymfam=\fiveym
  \fi
  \def\oldstyle{\fam\@ne\eighti}%
  \def\boldstyle{\fam\mibfam\eightmib}%
  \b@ls{10pt}\rm%
}

\def\ninepoint{
  \def\rm{\fam0\ninerm}%
  \textfont0=\ninerm \scriptfont0=\sixrm \scriptscriptfont0=\fiverm%
  \textfont1=\ninei  \scriptfont1=\sixi  \scriptscriptfont1=\fivei%
  \textfont2=\ninesy \scriptfont2=\sixsy \scriptscriptfont2=\fivesy%
  \textfont\itfam=\nineit\def\it{\fam\itfam\nineit}%
  \ifprod@font
    \scriptfont\itfam=\sixit
      \scriptscriptfont\itfam=\fiveit
  \else
    \scriptfont\itfam=\nineit
      \scriptscriptfont\itfam=\nineit
  \fi
  \textfont\bffam=\ninebf%
    \scriptfont\bffam=\sixbf%
      \scriptscriptfont\bffam=\fivebf%
  \def\bf{\fam\bffam\ninebf}%
  \textfont\slfam=\ninesl\def\sl{\fam\slfam\ninesl}%
  \ifprod@font
    \scriptfont\slfam=\sixsl
      \scriptscriptfont\slfam=\fivesl
  \else
    \scriptfont\slfam=\ninesl
      \scriptscriptfont\slfam=\ninesl
  \fi
  \textfont\ttfam=\ninett\def\tt{\fam\ttfam\ninett}%
  \ifprod@font
    \scriptfont\ttfam=\sixtt
      \scriptscriptfont\ttfam=\fivett
  \else
    \scriptfont\ttfam=\ninett
      \scriptscriptfont\ttfam=\ninett
  \fi
  \textfont\scfam=\ninecsc\def\sc{\fam\scfam\ninecsc}%
  \ifprod@font
    \scriptfont\scfam=\sixcsc
      \scriptscriptfont\scfam=\fivecsc
  \else
    \scriptfont\scfam=\ninecsc
      \scriptscriptfont\scfam=\ninecsc
  \fi
  \textfont\sffam=\ninesf\def\sf{\fam\sffam\ninesf}%
  \ifprod@font
    \scriptfont\sffam=\sixsf
      \scriptscriptfont\sffam=\fivesf
  \else
    \scriptfont\sffam=\ninesf
      \scriptscriptfont\sffam=\ninesf
  \fi
  \textfont\mibfam=\ninemib
    \scriptfont\mibfam=\sixmib
      \scriptscriptfont\mibfam=\fivemib
  \textfont\sybfam=\ninesyb
    \scriptfont\sybfam=\sixsyb
      \scriptscriptfont\sybfam=\fivesyb
  \ifprod@font
    \textfont\xmfam=\ninexm
      \scriptfont\xmfam=\sixxm
        \scriptscriptfont\xmfam=\fivexm
    \textfont\ymfam=\nineym
      \scriptfont\ymfam=\sixym
        \scriptscriptfont\ymfam=\fiveym
  \fi
  \def\oldstyle{\fam\@ne\ninei}%
  \def\boldstyle{\fam\mibfam\ninemib}%
  \b@ls{\TextLeading plus \Feathering}\rm%
}

\def\tenpoint{
  \def\rm{\fam0\tenrm}%
  \textfont0=\tenrm \scriptfont0=\sevenrm \scriptscriptfont0=\fiverm%
  \textfont1=\teni  \scriptfont1=\seveni  \scriptscriptfont1=\fivei%
  \textfont2=\tensy \scriptfont2=\sevensy \scriptscriptfont2=\fivesy%
  \textfont\itfam=\tenit\def\it{\fam\itfam\tenit}%
  \ifprod@font
    \scriptfont\itfam=\sevenit
      \scriptscriptfont\itfam=\fiveit
  \else
    \scriptfont\itfam=\tenit
      \scriptscriptfont\itfam=\tenit
  \fi
  \textfont\bffam=\tenbf%
    \scriptfont\bffam=\sevenbf%
      \scriptscriptfont\bffam=\fivebf%
  \def\bf{\fam\bffam\tenbf}%
  \textfont\slfam=\tensl\def\sl{\fam\slfam\tensl}%
  \ifprod@font
    \scriptfont\slfam=\sevensl
      \scriptscriptfont\slfam=\fivesl
  \else
    \scriptfont\slfam=\tensl
      \scriptscriptfont\slfam=\tensl
  \fi
  \textfont\ttfam=\tentt\def\tt{\fam\ttfam\tentt}%
  \ifprod@font
    \scriptfont\ttfam=\seventt
      \scriptscriptfont\ttfam=\fivett
  \else
    \scriptfont\ttfam=\tentt
      \scriptscriptfont\ttfam=\tentt
  \fi
  \textfont\scfam=\tencsc\def\sc{\fam\scfam\tencsc}%
  \ifprod@font
    \scriptfont\scfam=\sevencsc
      \scriptscriptfont\scfam=\fivecsc
  \else
    \scriptfont\scfam=\tencsc
      \scriptscriptfont\scfam=\tencsc
  \fi
  \textfont\sffam=\tensf\def\sf{\fam\sffam\tensf}%
  \ifprod@font
    \scriptfont\sffam=\sevensf
      \scriptscriptfont\sffam=\fivesf
  \else
    \scriptfont\sffam=\tensf
      \scriptscriptfont\sffam=\tensf
  \fi
  \textfont\mibfam=\tenmib
    \scriptfont\mibfam=\sevenmib
      \scriptscriptfont\mibfam=\fivemib
  \textfont\sybfam=\tensyb
    \scriptfont\sybfam=\sevensyb
      \scriptscriptfont\sybfam=\fivesyb
  \ifprod@font
    \textfont\xmfam=\tenxm
      \scriptfont\xmfam=\sevenxm
        \scriptscriptfont\xmfam=\fivexm
    \textfont\ymfam=\tenym
      \scriptfont\ymfam=\sevenym
        \scriptscriptfont\ymfam=\fiveym
  \fi
  \def\oldstyle{\fam\@ne\teni}%
  \def\boldstyle{\fam\mibfam\tenmib}%
  \b@ls{11pt}\rm%
}

\def\elevenpoint{
  \def\rm{\fam0\elevenrm}%
  \textfont0=\elevenrm \scriptfont0=\eightrm \scriptscriptfont0=\sixrm%
  \textfont1=\eleveni  \scriptfont1=\eighti  \scriptscriptfont1=\sixi%
  \textfont2=\elevensy \scriptfont2=\eightsy \scriptscriptfont2=\sixsy%
  \textfont\itfam=\elevenit\def\it{\fam\itfam\elevenit}%
  \ifprod@font
    \scriptfont\itfam=\eightit
      \scriptscriptfont\itfam=\sixit
  \else
    \scriptfont\itfam=\elevenit
      \scriptscriptfont\itfam=\elevenit
  \fi
  \textfont\bffam=\elevenbf%
    \scriptfont\bffam=\eightbf%
      \scriptscriptfont\bffam=\sixbf%
  \def\bf{\fam\bffam\elevenbf}%
  \textfont\slfam=\elevensl\def\sl{\fam\slfam\elevensl}%
  \ifprod@font
    \scriptfont\slfam=\eightsl
      \scriptscriptfont\slfam=\sixsl
  \else
    \scriptfont\slfam=\elevensl
      \scriptscriptfont\slfam=\elevensl
  \fi
  \textfont\ttfam=\eleventt\def\tt{\fam\ttfam\eleventt}%
  \ifprod@font
    \scriptfont\ttfam=\eighttt
      \scriptscriptfont\ttfam=\sixtt
  \else
    \scriptfont\ttfam=\eleventt
      \scriptscriptfont\ttfam=\eleventt
  \fi
  \textfont\scfam=\elevencsc\def\sc{\fam\scfam\elevencsc}%
  \ifprod@font
    \scriptfont\scfam=\eightcsc
      \scriptscriptfont\scfam=\sixcsc
  \else
    \scriptfont\scfam=\elevencsc
      \scriptscriptfont\scfam=\elevencsc
  \fi
  \textfont\sffam=\elevensf\def\sf{\fam\sffam\elevensf}%
  \ifprod@font
    \scriptfont\sffam=\eightsf
      \scriptscriptfont\sffam=\sixsf
  \else
    \scriptfont\sffam=\elevensf
      \scriptscriptfont\sffam=\elevensf
  \fi
  \textfont\mibfam=\elevenmib
    \scriptfont\mibfam=\eightmib
      \scriptscriptfont\mibfam=\sixmib
  \textfont\sybfam=\elevensyb
    \scriptfont\sybfam=\eightsyb
      \scriptscriptfont\sybfam=\sixsyb
  \ifprod@font
    \textfont\xmfam=\elevenxm
      \scriptfont\xmfam=\eightxm
       \scriptscriptfont\xmfam=\sixxm
    \textfont\ymfam=\elevenym
      \scriptfont\ymfam=\eightym
        \scriptscriptfont\ymfam=\sixym
   \fi
  \def\oldstyle{\fam\@ne\eleveni}%
  \def\boldstyle{\fam\mibfam\elevenmib}%
  \b@ls{13pt}\rm%
}

\def\fourteenpoint{
  \def\rm{\fam0\fourteenrm}%
  \textfont0\fourteenrm  \scriptfont0\tenrm  \scriptscriptfont0\sevenrm%
  \textfont1\fourteeni   \scriptfont1\teni   \scriptscriptfont1\seveni%
  \textfont2\fourteensy  \scriptfont2\tensy  \scriptscriptfont2\sevensy%
  \textfont\itfam=\fourteenit\def\it{\fam\itfam\fourteenit}%
  \ifprod@font
    \scriptfont\itfam=\tenit
      \scriptscriptfont\itfam=\sevenit
  \else
    \scriptfont\itfam=\fourteenit
      \scriptscriptfont\itfam=\fourteenit
  \fi
  \textfont\bffam=\fourteenbf%
    \scriptfont\bffam=\tenbf%
      \scriptscriptfont\bffam=\sevenbf%
  \def\bf{\fam\bffam\fourteenbf}%
  \textfont\slfam=\fourteensl\def\sl{\fam\slfam\fourteensl}%
  \ifprod@font
    \scriptfont\slfam=\tensl
      \scriptscriptfont\slfam=\sevensl
  \else
    \scriptfont\slfam=\fourteensl
      \scriptscriptfont\slfam=\fourteensl
  \fi
  \textfont\ttfam=\fourteentt\def\tt{\fam\ttfam\fourteentt}%
  \ifprod@font
    \scriptfont\ttfam=\tentt
      \scriptscriptfont\ttfam=\seventt
  \else
    \scriptfont\ttfam=\fourteentt
      \scriptscriptfont\ttfam=\fourteentt
  \fi
  \textfont\scfam=\fourteencsc\def\sc{\fam\scfam\fourteencsc}%
  \ifprod@font
    \scriptfont\scfam=\tencsc
      \scriptscriptfont\scfam=\sevencsc
  \else
    \scriptfont\scfam=\fourteencsc
      \scriptscriptfont\scfam=\fourteencsc
  \fi
  \textfont\sffam=\fourteensf\def\sf{\fam\sffam\fourteensf}%
  \ifprod@font
    \scriptfont\sffam=\tensf
      \scriptscriptfont\sffam=\sevensf
  \else
    \scriptfont\sffam=\fourteensf
      \scriptscriptfont\sffam=\fourteensf
  \fi
  \textfont\mibfam=\fourteenmib
    \scriptfont\mibfam=\tenmib
      \scriptscriptfont\mibfam=\sevenmib
  \textfont\sybfam=\fourteensyb
    \scriptfont\sybfam=\tensyb
      \scriptscriptfont\sybfam=\sevensyb
  \ifprod@font
    \textfont\xmfam=\fourteenxm
      \scriptfont\xmfam=\tenxm
        \scriptscriptfont\xmfam=\sevenxm
   \textfont\ymfam=\fourteenym
      \scriptfont\ymfam=\tenym
        \scriptscriptfont\ymfam=\sevenym
  \fi
  \def\oldstyle{\fam\@ne\fourteeni}%
  \def\boldstyle{\fam\mibfam\fourteenmib}%
  \b@ls{17pt}\rm%
}

\def\seventeenpoint{
  \def\rm{\fam0\seventeenrm}%
  \textfont0\seventeenrm  \scriptfont0\twelverm  \scriptscriptfont0\tenrm%
  \textfont1\seventeeni   \scriptfont1\twelvei   \scriptscriptfont1\teni%
  \textfont2\seventeensy  \scriptfont2\twelvesy  \scriptscriptfont2\tensy%
  \textfont\itfam=\seventeenit\def\it{\fam\itfam\seventeenit}%
  \ifprod@font
    \scriptfont\itfam=\twelveit
      \scriptscriptfont\itfam=\tenit
  \else
    \scriptfont\itfam=\seventeenit
      \scriptscriptfont\itfam=\seventeenit
  \fi
  \textfont\bffam=\seventeenbf%
    \scriptfont\bffam=\twelvebf%
      \scriptscriptfont\bffam=\tenbf%
  \def\bf{\fam\bffam\seventeenbf}%
  \textfont\slfam=\seventeensl\def\sl{\fam\slfam\seventeensl}%
  \ifprod@font
    \scriptfont\slfam=\twelvesl
      \scriptscriptfont\slfam=\tensl
  \else
    \scriptfont\slfam=\seventeensl
      \scriptscriptfont\slfam=\seventeensl
  \fi
  \textfont\ttfam=\seventeentt\def\tt{\fam\ttfam\seventeentt}%
  \ifprod@font
    \scriptfont\ttfam=\twelvett
      \scriptscriptfont\ttfam=\tentt
  \else
    \scriptfont\ttfam=\seventeentt
      \scriptscriptfont\ttfam=\seventeentt
  \fi
  \textfont\scfam=\seventeencsc\def\sc{\fam\scfam\seventeencsc}%
  \ifprod@font
    \scriptfont\scfam=\twelvecsc
      \scriptscriptfont\scfam=\tencsc
  \else
    \scriptfont\scfam=\seventeencsc
      \scriptscriptfont\scfam=\seventeencsc
  \fi
  \textfont\sffam=\seventeensf\def\sf{\fam\sffam\seventeensf}%
  \ifprod@font
    \scriptfont\sffam=\twelvesf
      \scriptscriptfont\sffam=\tensf
  \else
    \scriptfont\sffam=\seventeensf
      \scriptscriptfont\sffam=\seventeensf
  \fi
  \textfont\mibfam=\seventeenmib
    \scriptfont\mibfam=\twelvemib
      \scriptscriptfont\mibfam=\tenmib
  \textfont\sybfam=\seventeensyb
    \scriptfont\sybfam=\twelvesyb
      \scriptscriptfont\sybfam=\tensyb
  \ifprod@font
    \textfont\xmfam=\seventeenxm
      \scriptfont\xmfam=\twelvexm
        \scriptscriptfont\xmfam=\tenxm
    \textfont\ymfam=\seventeenym
      \scriptfont\ymfam=\twelveym
        \scriptscriptfont\ymfam=\tenym
  \fi
  \def\oldstyle{\fam\@ne\seventeeni}%
  \def\boldstyle{\fam\mibfam\seventeenmib}%
  \b@ls{20pt}\rm%
}

\lineskip=1pt      \normallineskip=\lineskip
\lineskiplimit=\z@ \normallineskiplimit=\lineskiplimit



\def\la{\mathrel{\mathchoice {\vcenter{\offinterlineskip\halign{\hfil
$\displaystyle##$\hfil\cr<\cr\sim\cr}}}
{\vcenter{\offinterlineskip\halign{\hfil$\textstyle##$\hfil\cr
<\cr\sim\cr}}}
{\vcenter{\offinterlineskip\halign{\hfil$\scriptstyle##$\hfil\cr
<\cr\sim\cr}}}
{\vcenter{\offinterlineskip\halign{\hfil$\scriptscriptstyle##$\hfil\cr
<\cr\sim\cr}}}}}

\def\ga{\mathrel{\mathchoice {\vcenter{\offinterlineskip\halign{\hfil
$\displaystyle##$\hfil\cr>\cr\sim\cr}}}
{\vcenter{\offinterlineskip\halign{\hfil$\textstyle##$\hfil\cr
>\cr\sim\cr}}}
{\vcenter{\offinterlineskip\halign{\hfil$\scriptstyle##$\hfil\cr
>\cr\sim\cr}}}
{\vcenter{\offinterlineskip\halign{\hfil$\scriptscriptstyle##$\hfil\cr
>\cr\sim\cr}}}}}

\def\getsto{\mathrel{\mathchoice {\vcenter{\offinterlineskip
\halign{\hfil
$\displaystyle##$\hfil\cr\gets\cr\to\cr}}}
{\vcenter{\offinterlineskip\halign{\hfil$\textstyle##$\hfil\cr\gets
\cr\to\cr}}}
{\vcenter{\offinterlineskip\halign{\hfil$\scriptstyle##$\hfil\cr\gets
\cr\to\cr}}}
{\vcenter{\offinterlineskip\halign{\hfil$\scriptscriptstyle##$\hfil\cr
\gets\cr\to\cr}}}}}

\def\lid{\mathrel{\mathchoice {\vcenter{\offinterlineskip\halign{\hfil
$\displaystyle##$\hfil\cr<\cr\noalign{\vskip1.2pt}=\cr}}}
{\vcenter{\offinterlineskip\halign{\hfil$\textstyle##$\hfil\cr<\cr
\noalign{\vskip1.2pt}=\cr}}}
{\vcenter{\offinterlineskip\halign{\hfil$\scriptstyle##$\hfil\cr<\cr
\noalign{\vskip1pt}=\cr}}}
{\vcenter{\offinterlineskip\halign{\hfil$\scriptscriptstyle##$\hfil\cr
<\cr
\noalign{\vskip0.9pt}=\cr}}}}}

\def\gid{\mathrel{\mathchoice {\vcenter{\offinterlineskip\halign{\hfil
$\displaystyle##$\hfil\cr>\cr\noalign{\vskip1.2pt}=\cr}}}
{\vcenter{\offinterlineskip\halign{\hfil$\textstyle##$\hfil\cr>\cr
\noalign{\vskip1.2pt}=\cr}}}
{\vcenter{\offinterlineskip\halign{\hfil$\scriptstyle##$\hfil\cr>\cr
\noalign{\vskip1pt}=\cr}}}
{\vcenter{\offinterlineskip\halign{\hfil$\scriptscriptstyle##$\hfil\cr
>\cr
\noalign{\vskip0.9pt}=\cr}}}}}

\def\grole{\mathrel{\mathchoice {\vcenter{\offinterlineskip\halign{\hfil
$\displaystyle##$\hfil\cr>\cr\noalign{\vskip-1.5pt}<\cr}}}
{\vcenter{\offinterlineskip\halign{\hfil$\textstyle##$\hfil\cr
>\cr\noalign{\vskip-1.5pt}<\cr}}}
{\vcenter{\offinterlineskip\halign{\hfil$\scriptstyle##$\hfil\cr
>\cr\noalign{\vskip-1pt}<\cr}}}
{\vcenter{\offinterlineskip\halign{\hfil$\scriptscriptstyle##$\hfil\cr
>\cr\noalign{\vskip-0.5pt}<\cr}}}}}

\def\leogr{\mathrel{\mathchoice {\vcenter{\offinterlineskip\halign{\hfil
$\displaystyle##$\hfil\cr<\cr\noalign{\vskip-1.5pt}>\cr}}}
{\vcenter{\offinterlineskip\halign{\hfil$\textstyle##$\hfil\cr
<\cr\noalign{\vskip-1.5pt}>\cr}}}
{\vcenter{\offinterlineskip\halign{\hfil$\scriptstyle##$\hfil\cr
<\cr\noalign{\vskip-1pt}>\cr}}}
{\vcenter{\offinterlineskip\halign{\hfil$\scriptscriptstyle##$\hfil\cr
<\cr\noalign{\vskip-0.5pt}>\cr}}}}}

\def\loa{\mathrel{\mathchoice {\vcenter{\offinterlineskip\halign{\hfil
$\displaystyle##$\hfil\cr<\cr\approx\cr}}}
{\vcenter{\offinterlineskip\halign{\hfil$\textstyle##$\hfil\cr
<\cr\approx\cr}}}
{\vcenter{\offinterlineskip\halign{\hfil$\scriptstyle##$\hfil\cr
<\cr\approx\cr}}}
{\vcenter{\offinterlineskip\halign{\hfil$\scriptscriptstyle##$\hfil\cr
<\cr\approx\cr}}}}}

\def\goa{\mathrel{\mathchoice {\vcenter{\offinterlineskip\halign{\hfil
$\displaystyle##$\hfil\cr>\cr\approx\cr}}}
{\vcenter{\offinterlineskip\halign{\hfil$\textstyle##$\hfil\cr
>\cr\approx\cr}}}
{\vcenter{\offinterlineskip\halign{\hfil$\scriptstyle##$\hfil\cr
>\cr\approx\cr}}}
{\vcenter{\offinterlineskip\halign{\hfil$\scriptscriptstyle##$\hfil\cr
>\cr\approx\cr}}}}}

\def\diameter{{\ifmmode\mathchoice
{\ooalign{\hfil\hbox{$\displaystyle/$}\hfil\crcr
{\hbox{$\displaystyle\mathchar"20D$}}}}
{\ooalign{\hfil\hbox{$\textstyle/$}\hfil\crcr
{\hbox{$\textstyle\mathchar"20D$}}}}
{\ooalign{\hfil\hbox{$\scriptstyle/$}\hfil\crcr
{\hbox{$\scriptstyle\mathchar"20D$}}}}
{\ooalign{\hfil\hbox{$\scriptscriptstyle/$}\hfil\crcr
{\hbox{$\scriptscriptstyle\mathchar"20D$}}}}
\else{\ooalign{\hfil/\hfil\crcr\mathhexbox20D}}%
\fi}}

\def\sq{\ifmmode\squareforqed\else{\unskip\nobreak\hfil
\penalty50\hskip1em\null\nobreak\hfil\squareforqed
\parfillskip=0pt\finalhyphendemerits=0\endgraf}\fi}
\def\squareforqed{\hbox{\rlap{$\sqcap$}$\sqcup$}}


\def\bbbc{{\mathchoice {\setbox0=\hbox{$\displaystyle\rm C$}\hbox{\hbox
to0pt{\kern0.4\wd0\vrule height0.9\ht0\hss}\box0}}
{\setbox0=\hbox{$\textstyle\rm C$}\hbox{\hbox
to0pt{\kern0.4\wd0\vrule height0.9\ht0\hss}\box0}}
{\setbox0=\hbox{$\scriptstyle\rm C$}\hbox{\hbox
to0pt{\kern0.4\wd0\vrule height0.9\ht0\hss}\box0}}
{\setbox0=\hbox{$\scriptscriptstyle\rm C$}\hbox{\hbox
to0pt{\kern0.4\wd0\vrule height0.9\ht0\hss}\box0}}}}
\def\bbbq{{\mathchoice {\setbox0=\hbox{$\displaystyle\rm
Q$}\hbox{\raise
0.15\ht0\hbox to0pt{\kern0.4\wd0\vrule height0.8\ht0\hss}\box0}}
{\setbox0=\hbox{$\textstyle\rm Q$}\hbox{\raise
0.15\ht0\hbox to0pt{\kern0.4\wd0\vrule height0.8\ht0\hss}\box0}}
{\setbox0=\hbox{$\scriptstyle\rm Q$}\hbox{\raise
0.15\ht0\hbox to0pt{\kern0.4\wd0\vrule height0.7\ht0\hss}\box0}}
{\setbox0=\hbox{$\scriptscriptstyle\rm Q$}\hbox{\raise
0.15\ht0\hbox to0pt{\kern0.4\wd0\vrule height0.7\ht0\hss}\box0}}}}
\def\bbbt{{\mathchoice {\setbox0=\hbox{$\displaystyle\rm
T$}\hbox{\hbox to0pt{\kern0.3\wd0\vrule height0.9\ht0\hss}\box0}}
{\setbox0=\hbox{$\textstyle\rm T$}\hbox{\hbox
to0pt{\kern0.3\wd0\vrule height0.9\ht0\hss}\box0}}
{\setbox0=\hbox{$\scriptstyle\rm T$}\hbox{\hbox
to0pt{\kern0.3\wd0\vrule height0.9\ht0\hss}\box0}}
{\setbox0=\hbox{$\scriptscriptstyle\rm T$}\hbox{\hbox
to0pt{\kern0.3\wd0\vrule height0.9\ht0\hss}\box0}}}}
\def\bbbs{{\mathchoice
{\setbox0=\hbox{$\displaystyle     \rm S$}\hbox{\raise0.5\ht0\hbox
to0pt{\kern0.35\wd0\vrule height0.45\ht0\hss}\hbox
to0pt{\kern0.55\wd0\vrule height0.5\ht0\hss}\box0}}
{\setbox0=\hbox{$\textstyle        \rm S$}\hbox{\raise0.5\ht0\hbox
to0pt{\kern0.35\wd0\vrule height0.45\ht0\hss}\hbox
to0pt{\kern0.55\wd0\vrule height0.5\ht0\hss}\box0}}
{\setbox0=\hbox{$\scriptstyle      \rm S$}\hbox{\raise0.5\ht0\hbox
to0pt{\kern0.35\wd0\vrule height0.45\ht0\hss}\raise0.05\ht0\hbox
to0pt{\kern0.5\wd0\vrule height0.45\ht0\hss}\box0}}
{\setbox0=\hbox{$\scriptscriptstyle\rm S$}\hbox{\raise0.5\ht0\hbox
to0pt{\kern0.4\wd0\vrule height0.45\ht0\hss}\raise0.05\ht0\hbox
to0pt{\kern0.55\wd0\vrule height0.45\ht0\hss}\box0}}}}
\def\bbbz{{\mathchoice {\hbox{$\sf\textstyle Z\kern-0.4em Z$}}
{\hbox{$\sf\textstyle Z\kern-0.4em Z$}}
{\hbox{$\sf\scriptstyle Z\kern-0.3em Z$}}
{\hbox{$\sf\scriptscriptstyle Z\kern-0.2em Z$}}}}


\ifprod@font
  \mathchardef\la="3\@xm2E
  \mathchardef\getsto="3\@xm1C
  \mathchardef\lid="3\@xm35
  \mathchardef\grole="3\@xm3F
  \mathchardef\loa="3\@xm2F
  \mathchardef\ga="3\@xm26
  \mathchardef\gid="3\@xm3D
  \mathchardef\leogr="3\@xm37
  \mathchardef\goa="3\@xm27
  \mathchardef\sq="0\@xm03
%
%
\def\diameter{{%
  \ifmmode
    \mathchoice
    {\ooalign{\hfil\hbox{$\displaystyle/$}\hfil\crcr
    {\lower.2ex\hbox{$\displaystyle\mathchar"20D$}}}}%
    {\ooalign{\hfil\hbox{$\textstyle/$}\hfil\crcr
    {\lower.2ex\hbox{$\textstyle\mathchar"20D$}}}}%
    {\ooalign{\hfil\hbox{$\scriptstyle/$}\hfil\crcr
    {\lower.1ex\hbox{$\scriptstyle\mathchar"20D$}}}}%
    {\ooalign{\hfil\hbox{$\scriptscriptstyle/$}\hfil\crcr
    {\lower.1ex\hbox{$\scriptscriptstyle\mathchar"20D$}}}}%
  \else
    {\ooalign{\hfil/\hfil\crcr\lower.2ex\hbox{\mathhexbox20D}}}%
  \fi
}}
%
%

\def\bbbc{{\Bbb{C}}}
\def\bbbq{{\Bbb{Q}}}
\def\bbbt{{\Bbb{T}}}
\def\bbbs{{\Bbb{S}}}
\def\bbbz{{\Bbb{Z}}}
\fi


\ifprod@font
\mathchardef\boxdot="2\@xm00
\mathchardef\boxplus="2\@xm01
\mathchardef\boxtimes="2\@xm02
\mathchardef\square="0\@xm03
\mathchardef\blacksquare="0\@xm04
\mathchardef\centerdot="2\@xm05
\mathchardef\lozenge="0\@xm06
\mathchardef\blacklozenge="0\@xm07
\mathchardef\circlearrowright="3\@xm08
\mathchardef\circlearrowleft="3\@xm09
\mathchardef\rightleftharpoons="3\@xm0A
\mathchardef\leftrightharpoons="3\@xm0B
\mathchardef\boxminus="2\@xm0C
\mathchardef\Vdash="3\@xm0D
\mathchardef\Vvdash="3\@xm0E
\mathchardef\vDash="3\@xm0F
\mathchardef\twoheadrightarrow="3\@xm10
\mathchardef\twoheadleftarrow="3\@xm11
\mathchardef\leftleftarrows="3\@xm12
\mathchardef\rightrightarrows="3\@xm13
\mathchardef\upuparrows="3\@xm14
\mathchardef\downdownarrows="3\@xm15
\mathchardef\upharpoonright="3\@xm16

\mathchardef\downharpoonright="3\@xm17
\mathchardef\upharpoonleft="3\@xm18
\mathchardef\downharpoonleft="3\@xm19
\mathchardef\rightarrowtail="3\@xm1A
\mathchardef\leftarrowtail="3\@xm1B
\mathchardef\leftrightarrows="3\@xm1C
\mathchardef\rightleftarrows="3\@xm1D
\mathchardef\Lsh="3\@xm1E
\mathchardef\Rsh="3\@xm1F
\mathchardef\rightsquigarrow="3\@xm20
\mathchardef\leftrightsquigarrow="3\@xm21
\mathchardef\looparrowleft="3\@xm22
\mathchardef\looparrowright="3\@xm23
\mathchardef\circeq="3\@xm24
\mathchardef\succsim="3\@xm25
\mathchardef\gtrsim="3\@xm26
\mathchardef\gtrapprox="3\@xm27
\mathchardef\multimap="3\@xm28
\mathchardef\therefore="3\@xm29
\mathchardef\because="3\@xm2A
\mathchardef\doteqdot="3\@xm2B

\mathchardef\triangleq="3\@xm2C
\mathchardef\precsim="3\@xm2D
\mathchardef\lesssim="3\@xm2E
\mathchardef\lessapprox="3\@xm2F
\mathchardef\eqslantless="3\@xm30
\mathchardef\eqslantgtr="3\@xm31
\mathchardef\curlyeqprec="3\@xm32
\mathchardef\curlyeqsucc="3\@xm33
\mathchardef\preccurlyeq="3\@xm34
\mathchardef\leqq="3\@xm35
\mathchardef\leqslant="3\@xm36
\mathchardef\lessgtr="3\@xm37
\mathchardef\backprime="0\@xm38
\mathchardef\risingdotseq="3\@xm3A
\mathchardef\fallingdotseq="3\@xm3B
\mathchardef\succcurlyeq="3\@xm3C
\mathchardef\geqq="3\@xm3D
\mathchardef\geqslant="3\@xm3E
\mathchardef\gtrless="3\@xm3F
\mathchardef\sqsubset="3\@xm40
\mathchardef\sqsupset="3\@xm41
\mathchardef\vartriangleright="3\@xm42
\mathchardef\vartriangleleft="3\@xm43
\mathchardef\trianglerighteq="3\@xm44
\mathchardef\trianglelefteq="3\@xm45
\mathchardef\bigstar="0\@xm46
\mathchardef\between="3\@xm47
\mathchardef\blacktriangledown="0\@xm48
\mathchardef\blacktriangleright="3\@xm49
\mathchardef\blacktriangleleft="3\@xm4A
\mathchardef\vartriangle="0\@xm4D
\mathchardef\blacktriangle="0\@xm4E
\mathchardef\triangledown="0\@xm4F
\mathchardef\eqcirc="3\@xm50
\mathchardef\lesseqgtr="3\@xm51
\mathchardef\gtreqless="3\@xm52
\mathchardef\lesseqqgtr="3\@xm53
\mathchardef\gtreqqless="3\@xm54
\mathchardef\Rrightarrow="3\@xm56
\mathchardef\Lleftarrow="3\@xm57
\mathchardef\veebar="2\@xm59
\mathchardef\barwedge="2\@xm5A
\mathchardef\doublebarwedge="2\@xm5B
\mathchardef\angle="0\@xm5C
\mathchardef\measuredangle="0\@xm5D
\mathchardef\sphericalangle="0\@xm5E
\mathchardef\varpropto="3\@xm5F
\mathchardef\smallsmile="3\@xm60
\mathchardef\smallfrown="3\@xm61
\mathchardef\Subset="3\@xm62
\mathchardef\Supset="3\@xm63
\mathchardef\Cup="2\@xm64

\mathchardef\Cap="2\@xm65

\mathchardef\curlywedge="2\@xm66
\mathchardef\curlyvee="2\@xm67
\mathchardef\leftthreetimes="2\@xm68
\mathchardef\rightthreetimes="2\@xm69
\mathchardef\subseteqq="3\@xm6A
\mathchardef\supseteqq="3\@xm6B
\mathchardef\bumpeq="3\@xm6C
\mathchardef\Bumpeq="3\@xm6D
\mathchardef\lll="3\@xm6E

\mathchardef\ggg="3\@xm6F

\mathchardef\circledS="0\@xm73
\mathchardef\pitchfork="3\@xm74
\mathchardef\dotplus="2\@xm75
\mathchardef\backsim="3\@xm76
\mathchardef\backsimeq="3\@xm77
\mathchardef\complement="0\@xm7B
\mathchardef\intercal="2\@xm7C
\mathchardef\circledcirc="2\@xm7D
\mathchardef\circledast="2\@xm7E
\mathchardef\circleddash="2\@xm7F
\def\ulcorner{\delimiter"4\@xm70\@xm70 }
\def\urcorner{\delimiter"5\@xm71\@xm71 }
\def\llcorner{\delimiter"4\@xm78\@xm78 }
\def\lrcorner{\delimiter"5\@xm79\@xm79 }
\def\yen{\mathhexbox\@xm55 }
\def\checkmark{\mathhexbox\@xm58 }
\def\circledR{\mathhexbox\@xm72 }
\def\maltese{\mathhexbox\@xm7A }
\mathchardef\lvertneqq="3\@ym00
\mathchardef\gvertneqq="3\@ym01
\mathchardef\nleq="3\@ym02
\mathchardef\ngeq="3\@ym03
\mathchardef\nless="3\@ym04
\mathchardef\ngtr="3\@ym05
\mathchardef\nprec="3\@ym06
\mathchardef\nsucc="3\@ym07
\mathchardef\lneqq="3\@ym08
\mathchardef\gneqq="3\@ym09
\mathchardef\nleqslant="3\@ym0A
\mathchardef\ngeqslant="3\@ym0B
\mathchardef\lneq="3\@ym0C
\mathchardef\gneq="3\@ym0D
\mathchardef\npreceq="3\@ym0E
\mathchardef\nsucceq="3\@ym0F
\mathchardef\precnsim="3\@ym10
\mathchardef\succnsim="3\@ym11
\mathchardef\lnsim="3\@ym12
\mathchardef\gnsim="3\@ym13
\mathchardef\nleqq="3\@ym14
\mathchardef\ngeqq="3\@ym15
\mathchardef\precneqq="3\@ym16
\mathchardef\succneqq="3\@ym17
\mathchardef\precnapprox="3\@ym18
\mathchardef\succnapprox="3\@ym19
\mathchardef\lnapprox="3\@ym1A
\mathchardef\gnapprox="3\@ym1B
\mathchardef\nsim="3\@ym1C
\mathchardef\ncong="3\@ym1D

\mathchardef\varsubsetneq="3\@ym20
\mathchardef\varsupsetneq="3\@ym21
\mathchardef\nsubseteqq="3\@ym22
\mathchardef\nsupseteqq="3\@ym23
\mathchardef\subsetneqq="3\@ym24
\mathchardef\supsetneqq="3\@ym25
\mathchardef\varsubsetneqq="3\@ym26
\mathchardef\varsupsetneqq="3\@ym27
\mathchardef\subsetneq="3\@ym28
\mathchardef\supsetneq="3\@ym29
\mathchardef\nsubseteq="3\@ym2A
\mathchardef\nsupseteq="3\@ym2B
\mathchardef\nparallel="3\@ym2C
\mathchardef\nmid="3\@ym2D
\mathchardef\nshortmid="3\@ym2E
\mathchardef\nshortparallel="3\@ym2F
\mathchardef\nvdash="3\@ym30
\mathchardef\nVdash="3\@ym31
\mathchardef\nvDash="3\@ym32
\mathchardef\nVDash="3\@ym33
\mathchardef\ntrianglerighteq="3\@ym34
\mathchardef\ntrianglelefteq="3\@ym35
\mathchardef\ntriangleleft="3\@ym36
\mathchardef\ntriangleright="3\@ym37
\mathchardef\nleftarrow="3\@ym38
\mathchardef\nrightarrow="3\@ym39
\mathchardef\nLeftarrow="3\@ym3A
\mathchardef\nRightarrow="3\@ym3B
\mathchardef\nLeftrightarrow="3\@ym3C
\mathchardef\nleftrightarrow="3\@ym3D
\mathchardef\divideontimes="2\@ym3E
\mathchardef\varnothing="0\@ym3F
\mathchardef\nexists="0\@ym40
\mathchardef\mho="0\@ym66
\mathchardef\eth="0\@ym67
\mathchardef\eqsim="3\@ym68
\mathchardef\beth="0\@ym69
\mathchardef\gimel="0\@ym6A
\mathchardef\daleth="0\@ym6B
\mathchardef\lessdot="3\@ym6C
\mathchardef\gtrdot="3\@ym6D
\mathchardef\ltimes="2\@ym6E
\mathchardef\rtimes="2\@ym6F
\mathchardef\shortmid="3\@ym70
\mathchardef\shortparallel="3\@ym71
\mathchardef\smallsetminus="2\@ym72
\mathchardef\thicksim="3\@ym73
\mathchardef\thickapprox="3\@ym74
\mathchardef\approxeq="3\@ym75
\mathchardef\succapprox="3\@ym76
\mathchardef\precapprox="3\@ym77
\mathchardef\curvearrowleft="3\@ym78
\mathchardef\curvearrowright="3\@ym79
\mathchardef\digamma="0\@ym7A
\mathchardef\varkappa="0\@ym7B
\mathchardef\hslash="0\@ym7D
\mathchardef\hbar="0\@ym7E
\mathchardef\backepsilon="3\@ym7F


\def\Bbb{\ifmmode\let\next\Bbb@\else
\def\next{\errmessage{Use \string\Bbb\space only in math mode}}\fi\next}
\def\Bbb@#1{{\Bbb@@{#1}}}
\def\Bbb@@#1{\fam\ymfam#1}
\fi


\def\Nulle{0} 
\def\Afe{1}   
\def\Hae{2}   
\def\Hbe{3}   
\def\Hce{4}   
\def\Hde{5}   


\newcount\LastMac       \LastMac=\Nulle

\newskip\half      \half=5.5pt plus 1.5pt minus 2.25pt
\newskip\one       \one=11pt plus 3pt minus 5.5pt
\newskip\onehalf   \onehalf=16.5pt plus 5.5pt minus 8.25pt
\newskip\two       \two=22pt plus 5.5pt minus 11pt

\def\Half{\addvspace{\half}}
\def\One{\addvspace{\one}}
\def\OneHalf{\addvspace{\onehalf}}
\def\Two{\addvspace{\two}}


\def\Raggedright{
  \rightskip=\z@ plus \hsize\relax
}

\def\Fullout{
  \rightskip=\z@\relax
}

\def\Hang#1#2{
  \hangindent=#1%
  \hangafter=#2\relax
}


\newif\ifsp@page
\def\pagestyle#1{\csname ps@#1\endcsname}
\def\thispagestyle#1{\global\sp@pagetrue\gdef\sp@type{#1}}

\def\ps@titlepage{%
  \def\@oddhead{\eightpoint\noindent \the\CatchLine
    \ifprod@font\else\qquad Printed\ \today\fi \hfil}%
  \let\@evenhead=\@oddhead
}

\def\ps@headings{%
  \def\@oddhead{\elevenpoint\it\noindent
    \hfill\the\RightHeader\hskip1.5em\rm\folio}%
  \def\@evenhead{\elevenpoint\noindent
    \folio\hskip1.5em\it\the\LeftHeader\hfill}%
}

\def\ps@plate{%
  \def\@oddhead{\eightpoint\noindent\plt@cap\hfil}%
  \def\@evenhead{\eightpoint\noindent\plt@cap\hfil}%
}



\def\title#1{
  \bgroup
    \vbox to 8pt{\vss}%
    \seventeenpoint
    \Raggedright
    \noindent \strut{\bf #1}\par
  \egroup
}

\def\author#1{
  \bgroup
    \ifnum\LastMac=\Afe \OneHalf\else \vskip 21pt\fi
    \fourteenpoint
    \Raggedright
    \noindent \strut #1\par
    \vskip 3pt%
  \egroup
}

\def\affiliation#1{
  \bgroup
    \vskip -4pt%
    \eightpoint
    \Raggedright
    \noindent \strut {\it #1}\par
  \egroup
  \LastMac=\Afe\relax
}

\def\acceptedline#1{
  \bgroup
    \Two
    \eightpoint
    \Raggedright
    \noindent \strut #1\par
  \egroup
}

\long\def\abstract#1{%
  \bgroup
    \vskip 20pt%
    \everypar{\Hang{11pc}{0}}%
    \noindent{\ninebf ABSTRACT}\par
    \tenpoint
    \Fullout
    \noindent #1\par
  \egroup
}

\long\def\keywords#1{
  \bgroup
    \Half
    \everypar{\Hang{11pc}{0}}%
    \tenpoint
    \Fullout
    \noindent\hbox{\bf Key words:}\ #1\par
  \egroup
}


\def\maketitle{%
  \EndOpening
  \ifsinglecol \else \MakePage\fi
}


\def\pageoffset#1#2{\hoffset=#1\relax\voffset=#2\relax}


\newif\ifAutoNumber \AutoNumberfalse
\newcount\Sec        
\newcount\SecSec
\newcount\SecSecSec

\Sec=\z@

\def\:{\let\@sptoken= } \:  
\def\:{\@xifnch} \expandafter\def\: {\futurelet\@tempc\@ifnch}

\def\@ifnextchar#1#2#3{%
  \let\@tempMACe #1%
  \def\@tempMACa{#2}%
  \def\@tempMACb{#3}%
  \futurelet \@tempMACc\@ifnch%
}

\def\@ifnch{%
\ifx \@tempMACc \@sptoken%
  \let\@tempMACd\@xifnch%
\else%
  \ifx \@tempMACc \@tempMACe%
    \let\@tempMACd\@tempMACa%
  \else%
    \let\@tempMACd\@tempMACb%
  \fi%
\fi%
\@tempMACd%
}

\def\@ifstar#1#2{\@ifnextchar *{\def\@tempMACa*{#1}\@tempMACa}{#2}}

\newskip\@tempskipb

\def\addvspace#1{%
  \ifvmode\else \endgraf\fi%
  \ifdim\lastskip=\z@%
    \vskip #1\relax%
  \else%
    \@tempskipb#1\relax\@xaddvskip%
  \fi%
}

\def\@xaddvskip{%
  \ifdim\lastskip<\@tempskipb%
    \vskip-\lastskip%
    \vskip\@tempskipb\relax%
  \else%
    \ifdim\@tempskipb<\z@%
      \ifdim\lastskip<\z@ \else%
        \advance\@tempskipb\lastskip%
        \vskip-\lastskip\vskip\@tempskipb%
      \fi%
    \fi%
  \fi%
}

\newskip\@tmpSKIP

\def\addpen#1{%
  \ifvmode
    \if@nobreak
    \else
      \ifdim\lastskip=\z@
        \penalty#1\relax
      \else
        \@tmpSKIP=\lastskip
        \vskip -\lastskip
        \penalty#1\vskip\@tmpSKIP
      \fi
    \fi
  \fi
}

\newcount\@clubpen   \@clubpen=\clubpenalty
\newif\if@nobreak    \@nobreakfalse

\def\@noafterindent{%
  \global\@nobreaktrue
  \everypar{\if@nobreak
              \global\@nobreakfalse
              \clubpenalty \@M
              {\setbox\z@\lastbox}%
              \LastMac=\Nulle\relax%
            \else
              \clubpenalty \@clubpen
              \everypar{}%
            \fi}
}

\newcount\gds@cbrk   \gds@cbrk=-300

\def\@nohdbrk{\interlinepenalty \@M\relax}

\let\@par=\par
\def\@restorepar{\def\par{\@par}}

\newif\if@endpe   \@endpefalse
 
\def\@doendpe{\@endpetrue \@nobreakfalse \LastMac=\Nulle\relax%
     \def\par{\@restorepar\everypar{}\par\@endpefalse}%
              \everypar{\setbox\z@\lastbox\everypar{}\@endpefalse}%
}

\def\section{\@ifstar{\@ssection}{\@section}}

\def\@section#1{
  \if@nobreak
    \everypar{}%
    \ifnum\LastMac=\Hae \addvspace{\half}\fi
  \else
    \addpen{\gds@cbrk}%
    \addvspace{\two}%
  \fi
  \bgroup
    \ninepoint\bf
    \Raggedright
    \ifAutoNumber
      \global\advance\Sec \@ne
      \noindent\@nohdbrk\number\Sec\hskip 1pc \uppercase{#1}\par
      \global\SecSec=\z@
    \else
      \noindent\@nohdbrk\uppercase{#1}\par
    \fi
  \egroup
  \nobreak
  \vskip\half
  \nobreak
  \@noafterindent
  \LastMac=\Hae\relax
}

\def\@ssection#1{
  \if@nobreak
    \everypar{}%
    \ifnum\LastMac=\Hae \addvspace{\half}\fi
  \else
    \addpen{\gds@cbrk}%
    \addvspace{\two}%
  \fi
  \bgroup
    \ninepoint\bf
    \Raggedright
    \noindent\@nohdbrk\uppercase{#1}\par
  \egroup
  \nobreak
  \vskip\half
  \nobreak
  \@noafterindent
  \LastMac=\Hae\relax
}

\def\subsection#1{
  \if@nobreak
    \everypar{}%
    \ifnum\LastMac=\Hae \addvspace{1pt plus 1pt minus .5pt}\fi
  \else
    \addpen{\gds@cbrk}%
    \addvspace{\onehalf}%
  \fi
  \bgroup
    \ninepoint\bf
    \Raggedright
    \ifAutoNumber
      \global\advance\SecSec \@ne
      \noindent\@nohdbrk\number\Sec.\number\SecSec \hskip 1pc\relax #1\par
      \global\SecSecSec=\z@
    \else
      \noindent\@nohdbrk #1\par
    \fi
  \egroup
  \nobreak
  \vskip\half
  \nobreak
  \@noafterindent
  \LastMac=\Hbe\relax
}

\def\subsubsection#1{
  \if@nobreak
    \everypar{}%
    \ifnum\LastMac=\Hbe \addvspace{1pt plus 1pt minus .5pt}\fi
  \else
    \addpen{\gds@cbrk}%
    \addvspace{\onehalf}%
  \fi
  \bgroup
    \ninepoint\it
    \Raggedright
    \ifAutoNumber
      \global\advance\SecSecSec \@ne
      \noindent\@nohdbrk\number\Sec.\number\SecSec.\number\SecSecSec
        \hskip 1pc\relax #1\par
    \else
      \noindent\@nohdbrk #1\par
    \fi
  \egroup
  \nobreak
  \vskip\half
  \nobreak
  \@noafterindent
  \LastMac=\Hce\relax
}

\def\paragraph#1{
  \if@nobreak
    \everypar{}%
  \else
    \addpen{\gds@cbrk}%
    \addvspace{\one}%
  \fi%
  \bgroup%
    \ninepoint\it
    \noindent #1\ \nobreak%
  \egroup
  \LastMac=\Hde\relax
  \ignorespaces
}


\let\tx=\relax 


\def\beginlist{%
  \par\if@nobreak \else\addvspace{\half}\fi%
  \bgroup%
    \ninepoint
    \let\item=\list@item%
}

\def\list@item{%
  \par\noindent\hskip 1em\relax%
  \ignorespaces%
}

\def\endlist{\par\egroup\addvspace{\half}\@doendpe}


\def\beginrefs{%
  \par
  \bgroup
    \eightpoint
    \Raggedright
    \let\bibitem=\bib@item
}

\def\bib@item{%
  \par\parindent=1.5em\Hang{1.5em}{1}%
  \everypar={\Hang{1.5em}{1}\ignorespaces}%
  \noindent\ignorespaces
}

\def\endrefs{\par\egroup\@doendpe}


\newtoks\CatchLine

\def\@journal{Mon.\ Not.\ R.\ Astron.\ Soc.\ }  
\def\@pubyear{1994}        
\def\@pagerange{000--000}  
\def\@volume{000}          
\def\@microfiche{}         %

\def\pubyear#1{\gdef\@pubyear{#1}\@makecatchline}
\def\pagerange#1{\gdef\@pagerange{#1}\@makecatchline}
\def\volume#1{\gdef\@volume{#1}\@makecatchline}
\def\microfiche#1{\gdef\@microfiche{and Microfiche\ #1}\@makecatchline}

\def\@makecatchline{%
  \global\CatchLine{%
    {\rm \@journal {\bf \@volume},\ \@pagerange\ (\@pubyear)\ \@microfiche}}%
}

\@makecatchline 

\newtoks\LeftHeader
\def\shortauthor#1{
  \global\LeftHeader{#1}%
}

\newtoks\RightHeader
\def\shorttitle#1{
  \global\RightHeader{#1}%
}

\def\PageHead{
  \begingroup
    \ifsp@page
      \csname ps@\sp@type\endcsname
      \global\sp@pagefalse
    \fi
    \ifodd\pageno
      \let\the@head=\@oddhead
    \else
      \let\the@head=\@evenhead
    \fi
    \vbox to \z@{\vskip-22.5\p@%
      \hbox to \PageWidth{\vbox to8.5\p@{}%
        \the@head
      }%
    \vss}%
  \endgroup
  \nointerlineskip
}

\def\today{%
  \number\day\space
  \ifcase\month\or January\or February\or March\or April\or May\or June\or
    July\or August\or September\or October\or November\or December\fi
  \space\number\year%
}

\def\PageFoot{} 

\def\authorcomment#1{%
  \gdef\PageFoot{%
    \nointerlineskip%
    \vbox to 22pt{\vfil%
      \hbox to \PageWidth{\elevenpoint\noindent \hfil #1 \hfil}}%
  }%
}


\newif\ifplate@page
\newbox\plt@box

\def\beginplatepage{%
  \let\plate=\plate@head
  \let\caption=\fig@caption
  \global\setbox\plt@box=\vbox\bgroup
  \TEMPDIMEN=\PageWidth 
  \hsize=\PageWidth\relax
}

\def\endplatepage{\par\egroup\global\plate@pagetrue}
\def\plate@head#1{\gdef\plt@cap{#1}}


\def\letters{%
  \gdef\folio{\ifnum\pageno<\z@ L\romannumeral-\pageno
    \else L\number\pageno \fi}%
}


\everydisplay{\displaysetup}

\newif\ifeqno
\newif\ifleqno

\def\displaysetup#1$${%
 \displaytest#1\eqno\eqno\displaytest
}

\def\displaytest#1\eqno#2\eqno#3\displaytest{%
 \if!#3!\ldisplaytest#1\leqno\leqno\ldisplaytest
 \else\eqnotrue\leqnofalse\def\eqn{#2}\def\eq{#1}\fi
 \generaldisplay$$}

\def\ldisplaytest#1\leqno#2\leqno#3\ldisplaytest{%
 \def\eq{#1}%
 \if!#3!\eqnofalse\else\eqnotrue\leqnotrue
  \def\eqn{#2}\fi}

\def\generaldisplay{%
\ifeqno \ifleqno 
   \hbox to \hsize{\noindent
     $\displaystyle\eq$\hfil$\displaystyle\eqn$}
  \else
    \hbox to \hsize{\noindent
     $\displaystyle\eq$\hfil$\displaystyle\eqn$}
  \fi
 \else
 \hbox to \hsize{\vbox{\noindent
  $\displaystyle\eq$\hfil}}
 \fi
}


\def\@notice{%
  \par\Two%
  \noindent{\b@ls{11pt}\ninerm This paper has been produced using the
    Blackwell Scientific Publications \TeX\ macros.\par}%
}

\outer\def\bye{\@notice\par\vfill\supereject\end}


\def\start@mess{%
  Monthly notices of the RAS journal style (\@typeface)\space
    v\@version,\space \@verdate.%
}

\everyjob{\Warn{\start@mess}}



\newif\if@debug \@debugfalse  

\def\Print#1{\if@debug\immediate\write16{#1}\else \fi}
\def\Warn#1{\immediate\write16{#1}}
\def\wlog#1{}

\newcount\Iteration 

\def\Single{0} \def\Double{1}                 
\def\Figure{0} \def\Table{1}                  

\def\InStack{0}  
\def\InZoneA{1}
\def\InZoneB{2}
\def\InZoneC{3}

\newcount\TEMPCOUNT 
\newdimen\TEMPDIMEN 
\newbox\TEMPBOX     
\newbox\VOIDBOX     

\newcount\LengthOfStack 
\newcount\MaxItems      
\newcount\StackPointer
\newcount\Point         
\newcount\NextFigure    
\newcount\NextTable     
\newcount\NextItem      

\newcount\StatusStack   
\newcount\NumStack      
\newcount\TypeStack     
\newcount\SpanStack     
\newcount\BoxStack      

\newcount\ItemSTATUS    
\newcount\ItemNUMBER    
\newcount\ItemTYPE      
\newcount\ItemSPAN      
\newbox\ItemBOX         
\newdimen\ItemSIZE      

\newdimen\PageHeight    
\newdimen\TextLeading   
\newdimen\Feathering    
\newcount\LinesPerPage  
\newdimen\ColumnWidth   
\newdimen\ColumnGap     
\newdimen\PageWidth     
\newdimen\BodgeHeight   
\newcount\Leading       

\newdimen\ZoneBSize  
\newdimen\TextSize   
\newbox\ZoneABOX     
\newbox\ZoneBBOX     
\newbox\ZoneCBOX     

\newif\ifFirstSingleItem
\newif\ifFirstZoneA
\newif\ifMakePageInComplete
\newif\ifMoreFigures \MoreFiguresfalse 
\newif\ifMoreTables  \MoreTablesfalse  

\newif\ifFigInZoneB 
\newif\ifFigInZoneC 
\newif\ifTabInZoneB 
\newif\ifTabInZoneC

\newif\ifZoneAFullPage

\newbox\MidBOX    
\newbox\LeftBOX
\newbox\RightBOX
\newbox\PageBOX   

\newif\ifLeftCOL  
\LeftCOLtrue

\newdimen\ZoneBAdjust

\newcount\ItemFits
\def\Yes{1}
\def\No{2}


\MaxItems=15
\NextFigure=\z@        
\NextTable=\@ne

\BodgeHeight=6pt
\TextLeading=11pt    
\Leading=11
\Feathering=\z@      
\LinesPerPage=61     
\topskip=\TextLeading
\ColumnWidth=20pc    
\ColumnGap=2pc       

\newskip\ItemSepamount  
\ItemSepamount=\TextLeading plus \TextLeading minus 4pt

\parskip=\z@ plus .1pt
\parindent=18pt
\widowpenalty=\z@
\clubpenalty=10000
\tolerance=1500
\hbadness=1500
\abovedisplayskip=6pt plus 2pt minus 2pt
\belowdisplayskip=6pt plus 2pt minus 2pt
\abovedisplayshortskip=6pt plus 2pt minus 2pt
\belowdisplayshortskip=6pt plus 2pt minus 2pt

\ninepoint 


\PageHeight=682pt

\PageWidth=2\ColumnWidth
\advance\PageWidth by \ColumnGap

\pagestyle{headings}




\newcount\DUMMY \StatusStack=\allocationnumber
\newcount\DUMMY \newcount\DUMMY \newcount\DUMMY 
\newcount\DUMMY \newcount\DUMMY \newcount\DUMMY 
\newcount\DUMMY \newcount\DUMMY \newcount\DUMMY
\newcount\DUMMY \newcount\DUMMY \newcount\DUMMY 
\newcount\DUMMY \newcount\DUMMY \newcount\DUMMY

\newcount\DUMMY \NumStack=\allocationnumber
\newcount\DUMMY \newcount\DUMMY \newcount\DUMMY 
\newcount\DUMMY \newcount\DUMMY \newcount\DUMMY 
\newcount\DUMMY \newcount\DUMMY \newcount\DUMMY 
\newcount\DUMMY \newcount\DUMMY \newcount\DUMMY 
\newcount\DUMMY \newcount\DUMMY \newcount\DUMMY

\newcount\DUMMY \TypeStack=\allocationnumber
\newcount\DUMMY \newcount\DUMMY \newcount\DUMMY 
\newcount\DUMMY \newcount\DUMMY \newcount\DUMMY 
\newcount\DUMMY \newcount\DUMMY \newcount\DUMMY 
\newcount\DUMMY \newcount\DUMMY \newcount\DUMMY 
\newcount\DUMMY \newcount\DUMMY \newcount\DUMMY

\newcount\DUMMY \SpanStack=\allocationnumber
\newcount\DUMMY \newcount\DUMMY \newcount\DUMMY 
\newcount\DUMMY \newcount\DUMMY \newcount\DUMMY 
\newcount\DUMMY \newcount\DUMMY \newcount\DUMMY 
\newcount\DUMMY \newcount\DUMMY \newcount\DUMMY 
\newcount\DUMMY \newcount\DUMMY \newcount\DUMMY

\newbox\DUMMY   \BoxStack=\allocationnumber
\newbox\DUMMY   \newbox\DUMMY \newbox\DUMMY 
\newbox\DUMMY   \newbox\DUMMY \newbox\DUMMY 
\newbox\DUMMY   \newbox\DUMMY \newbox\DUMMY 
\newbox\DUMMY   \newbox\DUMMY \newbox\DUMMY 
\newbox\DUMMY   \newbox\DUMMY \newbox\DUMMY

\def\wlog{\immediate\write\m@ne}


\def\GetItemAll#1{%
 \GetItemSTATUS{#1}
 \GetItemNUMBER{#1}
 \GetItemTYPE{#1}
 \GetItemSPAN{#1}
 \GetItemBOX{#1}
}

\def\GetItemSTATUS#1{%
 \Point=\StatusStack
 \advance\Point by #1
 \global\ItemSTATUS=\count\Point
}

\def\GetItemNUMBER#1{%
 \Point=\NumStack
 \advance\Point by #1
 \global\ItemNUMBER=\count\Point
}

\def\GetItemTYPE#1{%
 \Point=\TypeStack
 \advance\Point by #1
 \global\ItemTYPE=\count\Point
}

\def\GetItemSPAN#1{%
 \Point\SpanStack
 \advance\Point by #1
 \global\ItemSPAN=\count\Point
}

\def\GetItemBOX#1{%
 \Point=\BoxStack
 \advance\Point by #1
 \global\setbox\ItemBOX=\vbox{\copy\Point}
 \global\ItemSIZE=\ht\ItemBOX
 \global\advance\ItemSIZE by \dp\ItemBOX
 \TEMPCOUNT=\ItemSIZE
 \divide\TEMPCOUNT by \Leading
 \divide\TEMPCOUNT by 65536
 \advance\TEMPCOUNT \@ne
 \ItemSIZE=\TEMPCOUNT pt
 \global\multiply\ItemSIZE by \Leading
}


\def\JoinStack{%
 \ifnum\LengthOfStack=\MaxItems 
  \Warn{WARNING: Stack is full...some items will be lost!}
 \else
  \Point=\StatusStack
  \advance\Point by \LengthOfStack
  \global\count\Point=\ItemSTATUS
  \Point=\NumStack
  \advance\Point by \LengthOfStack
  \global\count\Point=\ItemNUMBER
  \Point=\TypeStack
  \advance\Point by \LengthOfStack
  \global\count\Point=\ItemTYPE
  \Point\SpanStack
  \advance\Point by \LengthOfStack
  \global\count\Point=\ItemSPAN
  \Point=\BoxStack
  \advance\Point by \LengthOfStack
  \global\setbox\Point=\vbox{\copy\ItemBOX}
  \global\advance\LengthOfStack \@ne
  \ifnum\ItemTYPE=\Figure 
   \global\MoreFigurestrue
  \else
   \global\MoreTablestrue
  \fi
 \fi
}


\def\LeaveStack#1{%
 {\Iteration=#1
 \loop
 \ifnum\Iteration<\LengthOfStack
  \advance\Iteration \@ne
  \GetItemSTATUS{\Iteration}
   \advance\Point by \m@ne
   \global\count\Point=\ItemSTATUS
  \GetItemNUMBER{\Iteration}
   \advance\Point by \m@ne
   \global\count\Point=\ItemNUMBER
  \GetItemTYPE{\Iteration}
   \advance\Point by \m@ne
   \global\count\Point=\ItemTYPE
  \GetItemSPAN{\Iteration}
   \advance\Point by \m@ne
   \global\count\Point=\ItemSPAN
  \GetItemBOX{\Iteration}
   \advance\Point by \m@ne
   \global\setbox\Point=\vbox{\copy\ItemBOX}
 \repeat}
 \global\advance\LengthOfStack by \m@ne
}


\newif\ifStackNotClean

\def\CleanStack{%
 \StackNotCleantrue
 {\Iteration=\z@
  \loop
   \ifStackNotClean
    \GetItemSTATUS{\Iteration}
    \ifnum\ItemSTATUS=\InStack
     \advance\Iteration \@ne
     \else
      \LeaveStack{\Iteration}
    \fi
   \ifnum\LengthOfStack<\Iteration
    \StackNotCleanfalse
   \fi
 \repeat}
}


\def\FindItem#1#2{%
 \global\StackPointer=\m@ne 
 {\Iteration=\z@
  \loop
  \ifnum\Iteration<\LengthOfStack
   \GetItemSTATUS{\Iteration}
   \ifnum\ItemSTATUS=\InStack
    \GetItemTYPE{\Iteration}
    \ifnum\ItemTYPE=#1
     \GetItemNUMBER{\Iteration}
     \ifnum\ItemNUMBER=#2
      \global\StackPointer=\Iteration
      \Iteration=\LengthOfStack 
     \fi
    \fi
   \fi
  \advance\Iteration \@ne
 \repeat}
}


\def\FindNext{%
 \global\StackPointer=\m@ne 
 {\Iteration=\z@
  \loop
  \ifnum\Iteration<\LengthOfStack
   \GetItemSTATUS{\Iteration}
   \ifnum\ItemSTATUS=\InStack
    \GetItemTYPE{\Iteration}
   \ifnum\ItemTYPE=\Figure
    \ifMoreFigures
      \global\NextItem=\Figure
      \global\StackPointer=\Iteration
      \Iteration=\LengthOfStack 
    \fi
   \fi
   \ifnum\ItemTYPE=\Table
    \ifMoreTables
      \global\NextItem=\Table
      \global\StackPointer=\Iteration
      \Iteration=\LengthOfStack 
    \fi
   \fi
  \fi
  \advance\Iteration \@ne
 \repeat}
}


\def\ChangeStatus#1#2{%
 \Point=\StatusStack
 \advance\Point by #1
 \global\count\Point=#2
}



\def\Zone{\InZoneA}

\ZoneBAdjust=\z@

\def\MakePage{
 \global\ZoneBSize=\PageHeight
 \global\TextSize=\ZoneBSize
 \global\ZoneAFullPagefalse
 \global\topskip=\TextLeading
 \MakePageInCompletetrue
 \MoreFigurestrue
 \MoreTablestrue
 \FigInZoneBfalse
 \FigInZoneCfalse
 \TabInZoneBfalse
 \TabInZoneCfalse
 \global\FirstSingleItemtrue
 \global\FirstZoneAtrue
 \global\setbox\ZoneABOX=\box\VOIDBOX
 \global\setbox\ZoneBBOX=\box\VOIDBOX
 \global\setbox\ZoneCBOX=\box\VOIDBOX
 \loop
  \ifMakePageInComplete
 \FindNext
 \ifnum\StackPointer=\m@ne
  \NextItem=\m@ne
  \MoreFiguresfalse
  \MoreTablesfalse
 \fi
 \ifnum\NextItem=\Figure
   \FindItem{\Figure}{\NextFigure}
   \ifnum\StackPointer=\m@ne \global\MoreFiguresfalse
   \else
    \GetItemSPAN{\StackPointer}
    \ifnum\ItemSPAN=\Single \def\Zone{\InZoneB}\relax
     \ifFigInZoneC \global\MoreFiguresfalse\fi
    \else
     \def\Zone{\InZoneA}
     \ifFigInZoneB \def\Zone{\InZoneC}\fi
    \fi
   \fi
   \ifMoreFigures\Print{}\FigureItems\fi
 \fi
\ifnum\NextItem=\Table
   \FindItem{\Table}{\NextTable}
   \ifnum\StackPointer=\m@ne \global\MoreTablesfalse
   \else
    \GetItemSPAN{\StackPointer}
    \ifnum\ItemSPAN=\Single\relax
     \ifTabInZoneC \global\MoreTablesfalse\fi
    \else
     \def\Zone{\InZoneA}
     \ifTabInZoneB \def\Zone{\InZoneC}\fi
    \fi
   \fi
   \ifMoreTables\Print{}\TableItems\fi
 \fi
   \MakePageInCompletefalse 
   \ifMoreFigures\MakePageInCompletetrue\fi
   \ifMoreTables\MakePageInCompletetrue\fi
 \repeat
 \ifZoneAFullPage
  \global\TextSize=\z@
  \global\ZoneBSize=\z@
  \global\vsize=\z@\relax
  \global\topskip=\z@\relax
  \vbox to \z@{\vss}
  \eject
 \else
 \global\advance\ZoneBSize by -\ZoneBAdjust
 \global\vsize=\ZoneBSize
 \global\hsize=\ColumnWidth
 \global\ZoneBAdjust=\z@
 \ifdim\TextSize<23pt
 \Warn{}
 \Warn{* Making column fall short: TextSize=\the\TextSize *}
 \vskip-\lastskip\eject\fi
 \fi
}

\def\MakeRightCol{
 \global\TextSize=\ZoneBSize
 \MakePageInCompletetrue
 \MoreFigurestrue
 \MoreTablestrue
 \global\FirstSingleItemtrue
 \global\setbox\ZoneBBOX=\box\VOIDBOX
 \def\Zone{\InZoneB}
 \loop
  \ifMakePageInComplete
 \FindNext
 \ifnum\StackPointer=\m@ne
  \NextItem=\m@ne
  \MoreFiguresfalse
  \MoreTablesfalse
 \fi
 \ifnum\NextItem=\Figure
   \FindItem{\Figure}{\NextFigure}
   \ifnum\StackPointer=\m@ne \MoreFiguresfalse
   \else
    \GetItemSPAN{\StackPointer}
    \ifnum\ItemSPAN=\Double\relax
     \MoreFiguresfalse\fi
   \fi
   \ifMoreFigures\Print{}\FigureItems\fi
 \fi
 \ifnum\NextItem=\Table
   \FindItem{\Table}{\NextTable}
   \ifnum\StackPointer=\m@ne \MoreTablesfalse
   \else
    \GetItemSPAN{\StackPointer}
    \ifnum\ItemSPAN=\Double\relax
     \MoreTablesfalse\fi
   \fi
   \ifMoreTables\Print{}\TableItems\fi
 \fi
   \MakePageInCompletefalse 
   \ifMoreFigures\MakePageInCompletetrue\fi
   \ifMoreTables\MakePageInCompletetrue\fi
 \repeat
 \ifZoneAFullPage
  \global\TextSize=\z@
  \global\ZoneBSize=\z@
  \global\vsize=\z@\relax
  \global\topskip=\z@\relax
  \vbox to \z@{\vss}
  \eject
 \else
 \global\vsize=\ZoneBSize
 \global\hsize=\ColumnWidth
 \ifdim\TextSize<23pt
 \Warn{}
 \Warn{* Making column fall short: TextSize=\the\TextSize *}
 \vskip-\lastskip\eject\fi
\fi
}

\def\FigureItems{
 \Print{Considering...}
 \ShowItem{\StackPointer}
 \GetItemBOX{\StackPointer} 
 \GetItemSPAN{\StackPointer}
  \CheckFitInZone 
  \ifnum\ItemFits=\Yes
   \ifnum\ItemSPAN=\Single
     \ChangeStatus{\StackPointer}{\InZoneB} 
     \global\FigInZoneBtrue
     \ifFirstSingleItem
      \hbox{}\vskip-\BodgeHeight
     \global\advance\ItemSIZE by \TextLeading
     \fi
     \unvbox\ItemBOX\ItemSep
     \global\FirstSingleItemfalse
     \global\advance\TextSize by -\ItemSIZE
     \global\advance\TextSize by -\TextLeading
   \else
    \ifFirstZoneA
     \global\advance\ItemSIZE by \TextLeading
     \global\FirstZoneAfalse\fi
    \global\advance\TextSize by -\ItemSIZE
    \global\advance\TextSize by -\TextLeading
    \global\advance\ZoneBSize by -\ItemSIZE
    \global\advance\ZoneBSize by -\TextLeading
    \ifFigInZoneB\relax
     \else
     \ifdim\TextSize<3\TextLeading
     \global\ZoneAFullPagetrue
     \fi
    \fi
    \ChangeStatus{\StackPointer}{\Zone}
    \ifnum\Zone=\InZoneC \global\FigInZoneCtrue\fi
  \fi
   \Print{TextSize=\the\TextSize}
   \Print{ZoneBSize=\the\ZoneBSize}
  \global\advance\NextFigure \@ne
   \Print{This figure has been placed.}
  \else
   \Print{No space available for this figure...holding over.}
   \Print{}
   \global\MoreFiguresfalse
  \fi
}

\def\TableItems{
 \Print{Considering...}
 \ShowItem{\StackPointer}
 \GetItemBOX{\StackPointer} 
 \GetItemSPAN{\StackPointer}
  \CheckFitInZone 
  \ifnum\ItemFits=\Yes
   \ifnum\ItemSPAN=\Single
    \ChangeStatus{\StackPointer}{\InZoneB}
     \global\TabInZoneBtrue
     \ifFirstSingleItem
      \hbox{}\vskip-\BodgeHeight
     \global\advance\ItemSIZE by \TextLeading
     \fi
     \unvbox\ItemBOX\ItemSep
     \global\FirstSingleItemfalse
     \global\advance\TextSize by -\ItemSIZE
     \global\advance\TextSize by -\TextLeading
   \else
    \ifFirstZoneA
    \global\advance\ItemSIZE by \TextLeading
    \global\FirstZoneAfalse\fi
    \global\advance\TextSize by -\ItemSIZE
    \global\advance\TextSize by -\TextLeading
    \global\advance\ZoneBSize by -\ItemSIZE
    \global\advance\ZoneBSize by -\TextLeading
    \ifFigInZoneB\relax
     \else
     \ifdim\TextSize<3\TextLeading
     \global\ZoneAFullPagetrue
     \fi
    \fi
    \ChangeStatus{\StackPointer}{\Zone}
    \ifnum\Zone=\InZoneC \global\TabInZoneCtrue\fi
   \fi
  \global\advance\NextTable \@ne
   \Print{This table has been placed.}
  \else
  \Print{No space available for this table...holding over.}
   \Print{}
   \global\MoreTablesfalse
  \fi
}


\def\CheckFitInZone{%
{\advance\TextSize by -\ItemSIZE
 \advance\TextSize by -\TextLeading
 \ifFirstSingleItem
  \advance\TextSize by \TextLeading
 \fi
 \ifnum\Zone=\InZoneA\relax
  \else \advance\TextSize by -\ZoneBAdjust
 \fi
 \ifdim\TextSize<3\TextLeading \global\ItemFits=\No
 \else \global\ItemFits=\Yes\fi}
}

\def\BeginOpening{%
  \thispagestyle{titlepage}%
  \global\setbox\ItemBOX=\vbox\bgroup%
    \hsize=\PageWidth%
    \hrule height \z@
    \ifsinglecol\vskip 6pt\fi 
}

\let\begintopmatter=\BeginOpening  

\def\EndOpening{%
  \One
  \egroup
  \ifsinglecol
    \box\ItemBOX%
    \vskip\TextLeading plus 2\TextLeading
    \@noafterindent
  \else
    \ItemNUMBER=\z@%
    \ItemTYPE=\Figure
    \ItemSPAN=\Double
    \ItemSTATUS=\InStack
    \JoinStack
  \fi
}


\newif\if@here  \@herefalse

\def\no@float{\global\@heretrue}
\let\nofloat=\relax 

\def\beginfigure{%
  \@ifstar{\global\@dfloattrue \@bfigure}{\global\@dfloatfalse \@bfigure}%
}

\def\@bfigure#1{%
  \par
  \if@dfloat
    \ItemSPAN=\Double
    \TEMPDIMEN=\PageWidth
  \else
    \ItemSPAN=\Single
    \TEMPDIMEN=\ColumnWidth
  \fi
  \ifsinglecol
    \TEMPDIMEN=\PageWidth
  \else
    \ItemSTATUS=\InStack
    \ItemNUMBER=#1%
    \ItemTYPE=\Figure
  \fi
  \bgroup
    \hsize=\TEMPDIMEN
    \global\setbox\ItemBOX=\vbox\bgroup
      \eightpoint\nostb@ls{10pt}%
      \let\caption=\fig@caption
      \ifsinglecol \let\nofloat=\no@float\fi
}

\def\fig@caption#1{%
  \vskip 5.5pt plus 6pt%
  \bgroup 
    \eightpoint\nostb@ls{10pt}%
    \setbox\TEMPBOX=\hbox{#1}%
    \ifdim\wd\TEMPBOX>\TEMPDIMEN
      \noindent \unhbox\TEMPBOX\par
    \else
      \hbox to \hsize{\hfil\unhbox\TEMPBOX\hfil}%
    \fi
  \egroup
}

\def\endfigure{%
  \par\egroup 
  \egroup
  \ifsinglecol
    \if@here \midinsert\global\@herefalse\else \topinsert\fi
      \unvbox\ItemBOX
    \endinsert
  \else
    \JoinStack
    \Print{Processing source for figure \the\ItemNUMBER}%
  \fi
}


\newbox\tab@cap@box
\def\tab@caption#1{\global\setbox\tab@cap@box=\hbox{#1\par}}

\newtoks\tab@txt@toks
\long\def\tab@txt#1{\global\tab@txt@toks={#1}\global\table@txttrue}

\newif\iftable@txt  \table@txtfalse
\newif\if@dfloat    \@dfloatfalse

\def\begintable{%
  \@ifstar{\global\@dfloattrue \@btable}{\global\@dfloatfalse \@btable}%
}

\def\@btable#1{%
  \par
  \if@dfloat
    \ItemSPAN=\Double
    \TEMPDIMEN=\PageWidth
  \else
    \ItemSPAN=\Single
    \TEMPDIMEN=\ColumnWidth
  \fi
  \ifsinglecol
    \TEMPDIMEN=\PageWidth
  \else
    \ItemSTATUS=\InStack
    \ItemNUMBER=#1%
    \ItemTYPE=\Table
  \fi
  \bgroup
    \eightpoint\nostb@ls{10pt}%
    \global\setbox\ItemBOX=\vbox\bgroup
      \let\caption=\tab@caption
      \let\tabletext=\tab@txt
      \ifsinglecol \let\nofloat=\no@float\fi
}

\def\endtable{%
  \par\egroup 
  \egroup
  \setbox\TEMPBOX=\hbox to \TEMPDIMEN{%
    \hss
    \vbox{%
      \hsize=\wd\ItemBOX
      \ifvoid\tab@cap@box
      \else
        \noindent\unhbox\tab@cap@box
        \vskip 5.5pt plus 6pt%
      \fi
      \box\ItemBOX
      \iftable@txt
        \vskip 10pt%
        \eightpoint\nostb@ls{10pt}%
        \noindent\the\tab@txt@toks
        \global\table@txtfalse
      \fi
    }%
    \hss
  }%
  \ifsinglecol
    \if@here \midinsert\global\@herefalse\else \topinsert\fi
      \box\TEMPBOX
    \endinsert
  \else
    \global\setbox\ItemBOX=\box\TEMPBOX
    \JoinStack
    \Print{Processing source for table \the\ItemNUMBER}%
  \fi
}

\def\UnloadZoneA{%
\FirstZoneAtrue
 \Iteration=\z@
  \loop
   \ifnum\Iteration<\LengthOfStack
    \GetItemSTATUS{\Iteration}
    \ifnum\ItemSTATUS=\InZoneA
     \GetItemBOX{\Iteration}
     \ifFirstZoneA \vbox to \BodgeHeight{\vfil}%
     \FirstZoneAfalse\fi
     \unvbox\ItemBOX\ItemSep
     \LeaveStack{\Iteration}
     \else
     \advance\Iteration \@ne
   \fi
 \repeat
}

\def\UnloadZoneC{%
\Iteration=\z@
  \loop
   \ifnum\Iteration<\LengthOfStack
    \GetItemSTATUS{\Iteration}
    \ifnum\ItemSTATUS=\InZoneC
     \GetItemBOX{\Iteration}
     \ItemSep\unvbox\ItemBOX
     \LeaveStack{\Iteration}
     \else
     \advance\Iteration \@ne
   \fi
 \repeat
}


\def\ShowItem#1{
  {\GetItemAll{#1}
  \Print{\the#1:
  {TYPE=\ifnum\ItemTYPE=\Figure Figure\else Table\fi}
  {NUMBER=\the\ItemNUMBER}
  {SPAN=\ifnum\ItemSPAN=\Single Single\else Double\fi}
  {SIZE=\the\ItemSIZE}}}
}

\def\ShowStack{%
 \Print{}
 \Print{LengthOfStack = \the\LengthOfStack}
 \ifnum\LengthOfStack=\z@ \Print{Stack is empty}\fi
 \Iteration=\z@
 \loop
 \ifnum\Iteration<\LengthOfStack
  \ShowItem{\Iteration}
  \advance\Iteration \@ne
 \repeat
}

\def\B#1#2{%
\hbox{\vrule\kern-0.4pt\vbox to #2{%
\hrule width #1\vfill\hrule}\kern-0.4pt\vrule}
}


\newif\ifsinglecol   \singlecolfalse

\def\onecolumn{%
  \global\output={\singlecoloutput}%
  \global\hsize=\PageWidth
  \global\vsize=\PageHeight
  \global\ColumnWidth=\hsize
  \global\TextLeading=12pt
  \global\Leading=12
  \global\singlecoltrue
  \global\let\onecolumn=\relax
  \global\let\footnote=\sing@footnote
  \global\let\vfootnote=\sing@vfootnote
  \ninepoint 
  \message{(Single column)}%
}

\def\singlecoloutput{%
  \shipout\vbox{\PageHead\pagebody\PageFoot}%
  \advancepageno
  \ifplate@page
    \shipout\vbox{%
      \sp@pagetrue
      \def\sp@type{plate}%
      \global\plate@pagefalse
      \PageHead\vbox to \PageHeight{\unvbox\plt@box\vfil}\PageFoot%
    }%
    \message{[plate]}%
    \advancepageno
  \fi
  \ifnum\outputpenalty>-\@MM \else\dosupereject\fi%
}

\def\ItemSep{\vskip\ItemSepamount\relax}

\def\ItemSepbreak{\par\ifdim\lastskip<\ItemSepamount
  \removelastskip\penalty-200\ItemSep\fi%
}


\let\@@endinsert=\endinsert 

\def\endinsert{\egroup 
  \if@mid \dimen@\ht\z@ \advance\dimen@\dp\z@ \advance\dimen@12\p@
    \advance\dimen@\pagetotal \advance\dimen@-\pageshrink
    \ifdim\dimen@>\pagegoal\@midfalse\p@gefalse\fi\fi
  \if@mid \ItemSep\box\z@\ItemSepbreak
  \else\insert\topins{\penalty100 
    \splittopskip\z@skip
    \splitmaxdepth\maxdimen \floatingpenalty\z@
    \ifp@ge \dimen@\dp\z@
    \vbox to\vsize{\unvbox\z@\kern-\dimen@}
    \else \box\z@\nobreak\ItemSep\fi}\fi\endgroup%
}


\def\gobbleone#1{}
\def\gobbletwo#1#2{}
\let\footnote=\gobbletwo 
\let\vfootnote=\gobbleone

\def\sing@footnote#1{\let\@sf\empty 
  \ifhmode\edef\@sf{\spacefactor\the\spacefactor}\/\fi
  \hbox{$^{\hbox{\eightpoint #1}}$}\@sf\sing@vfootnote{#1}%
}

\def\sing@vfootnote#1{\insert\footins\bgroup\eightpoint\b@ls{9pt}%
  \interlinepenalty\interfootnotelinepenalty
  \splittopskip\ht\strutbox 
  \splitmaxdepth\dp\strutbox \floatingpenalty\@MM
  \leftskip\z@skip \rightskip\z@skip \spaceskip\z@skip \xspaceskip\z@skip
  \noindent $^{\scriptstyle\hbox{#1}}$\hskip 4pt%
    \footstrut\futurelet\next\fo@t%
}

\def\footnoterule{\kern-3\p@ \hrule height \z@ \kern 3\p@}

\skip\footins=19.5pt plus 12pt minus 1pt
\count\footins=1000
\dimen\footins=\maxdimen


\def\landscape{%
  \global\TEMPDIMEN=\PageWidth
  \global\PageWidth=\PageHeight
  \global\PageHeight=\TEMPDIMEN
  \global\let\landscape=\relax
  \onecolumn
  \message{(landscape)}%
  \raggedbottom
}


\output{%
  \ifLeftCOL
    \global\setbox\LeftBOX=\vbox to \ZoneBSize{\box255\unvbox\ZoneBBOX}%
    \global\LeftCOLfalse
    \MakeRightCol
  \else
    \setbox\RightBOX=\vbox to \ZoneBSize{\box255\unvbox\ZoneBBOX}%
    \setbox\MidBOX=\hbox{\box\LeftBOX\hskip\ColumnGap\box\RightBOX}%
    \setbox\PageBOX=\vbox to \PageHeight{%
      \UnloadZoneA\box\MidBOX\UnloadZoneC}%
    \shipout\vbox{\PageHead\box\PageBOX\PageFoot}%
    \advancepageno
    \ifplate@page
      \shipout\vbox{%
        \sp@pagetrue
        \def\sp@type{plate}%
        \global\plate@pagefalse
        \PageHead\vbox to \PageHeight{\unvbox\plt@box\vfil}\PageFoot%
      }%
      \message{[plate]}%
      \advancepageno
    \fi
    \global\LeftCOLtrue
    \CleanStack
    \MakePage
  \fi
}


\Warn{\start@mess}


\catcode `\@=12 




\newif\ifprintcomments
 
 
\printcommentstrue


\def\etal{et al.~}


\def\Msun{{\rm\,M_\odot}}
\def\Lsun{{\rm\,L_\odot}}
\def\kms{{\rm\,km\,s^{-1}}}



\def\deg{^{\circ}}


\def\spose#1{\hbox to 0pt{#1\hss}}
\def\lta{\mathrel{\spose{\lower 3pt\hbox{$\sim$}}
    \raise 2.0pt\hbox{$<$}}}
\def\gta{\mathrel{\spose{\lower 3pt\hbox{$\sim$}}
    \raise 2.0pt\hbox{$>$}}}


\newdimen\hssize
\hssize=8.4truecm
\newdimen\hdsize
\hdsize=17.7truecm


\def\today{\ifcase\month\or
 January\or February\or March\or April\or May\or June\or
 July\or August\or September\or October\or November\or December\fi
 \space\number\day, \number\year}
 

\newcount\eqnumber
\eqnumber=1
 
\def\new{\hbox{(\the\eqnumber )}\global\advance\eqnumber by 1}
 
\def\first{\hbox{(\the\eqnumber a)}\global\advance\eqnumber by 1}
\def\last#1{\advance\eqnumber by -1 \hbox{(\the\eqnumber#1)}\advance
     \eqnumber by 1}
 
\def\ref#1{\advance\eqnumber by -#1 \the\eqnumber
     \advance\eqnumber by #1}
 
\def\nref#1{\advance\eqnumber by -#1 \the\eqnumber
     \advance\eqnumber by #1}

\def\eqnam#1{\xdef#1{\the\eqnumber}}



\pageoffset{-0.85truecm}{-1.05truecm}



\pagerange{}
\pubyear{version: \today; {\tt DO NOT DISTRIBUTE}}
\volume{}

\begintopmatter

\title{Nuclear stellar discs in early-type galaxies --- 
\hfill\vskip0.001truecm
{\noindent II. Photometric properties}}

\author{Cecilia Scorza$^{1,2}$ and Frank C. van den Bosch$^{3,4}$}

\affiliation{$^1$ Landessternwarte, K\"onigstuhl, D-69117 Heidelberg, Germany}
\medskip
\affiliation{$^2$ Centro de Investigaciones de Astronomia CIDA, Merida, 
Venezuela}
\medskip
\affiliation{$^3$ Department of Astronomy, University of Washington, Seattle, 
             WA 98195, USA}
\medskip
\affiliation{$^4$ Hubble Fellow}

\shortauthor{C. Scorza and F.C. van den Bosch}

\shorttitle{%
Nuclear stellar discs in early-type galaxies}


\abstract{%
  Hubble Space Telescope images of two early-type galaxies
  harboring both nuclear and outer stellar discs are studied in
  detail. By means of a photometric decomposition, the images of
  NGC~4342 and NGC~4570 are analyzed and the photometric properties of
  the nuclear discs investigated.  We find a continuity of properties
  in the parameter space defined by the central surface brightness
  $\mu_0$ and the scalelength $R_d$ of discs in spirals, S0s and
  embedded discs in ellipticals, in the sense that the nuclear discs
  extend the observed disc properties even further towards smaller
  scalelengths and brighter central surface brightnesses. When
  including the nuclear discs, disc properties span more than four
  orders of magnitude in both scalelength and central surface
  brightness.  The nuclear discs studied here are the smallest and
  brightest stellar discs known, and as such, they are as extreme in
  their photometric properties as Malin I, when compared to typical
  galactic discs that obey Freeman's law. We discuss a possible
  formation scenario in which the double-disc structure observed in
  these galaxies has been shaped by now dissolved bars.  Based on the
  fact that the black holes known to exist in some of these galaxies
  have masses comparable to those of the nuclear discs, we explore a
  possible link between the black holes and the nuclear discs.
}

\keywords{%
  galaxies: disk -- galaxies: elliptical and lenticular, cD --
  galaxies: photometry -- galaxies: individual: NGC~4342, NGC~4570
}

\maketitle  


\section{1 Introduction}

The dichotomy amongst early-type galaxies inferred some years ago from
ground-based observations (Bender 1988; Bender \etal 1989; Nieto
\etal 1991) has recently been confirmed and strengthened with high
spatial resolution images obtained with the Hubble Space Telescope
(HST).  These images show that early-type galaxies can be divided in
two classes according to the shape of the surface brightness profiles
in the central regions: the bright pressured supported boxy systems
show central luminosity profiles which can be well fit by a double
power-law (they show a clear break); the low luminosity discy galaxies
have brightness profiles that lack a clear break and keep rising
steeply down to the smallest scales accessible by the HST (e.g.,
Ferrarese \etal 1994; Lauer \etal 1995; Gebhardt \etal 1996). Lauer
\etal (1995) refer to the bright ellipticals with a clear inner
break-radius as ``core galaxies'' and to the low luminosity
ellipticals with single power-law luminosity profiles as ``power-law''
galaxies. They were termed Type~I and Type~II respectively by Jaffe
\etal (1994).

The high resolution HST images revealed the presence of small nuclear
stellar discs in a number of power-law early-type galaxies (van den
Bosch \etal 1994; Lauer \etal 1995; van den Bosch, Jaffe \& van der
Marel 1998).  These galaxies are in general composite systems,
consisting of a bulge and a large or intermediate kpc-scale disc.  A
well known example is the Sombrero galaxy, in which besides the outer
kpc-scale disc, a bright nuclear disc had already been discovered from
ground-based observations (Burkhead 1986, 1991; Kormendy 1988;
Emsellem \etal 1996). Most of the galaxies that harbor a nuclear disc
are either S0s or discy ellipticals. The latter harbor fully-embedded
kpc-scale stellar discs (Carter 1987; Bender 1990; Nieto \etal 1991;
Rix \& White 1990, 1992; Scorza \& Bender 1990, 1995), which are
smaller and brighter than discs in S0s and spirals (Scorza \& Bender
1995, 1996), but significantly larger than the nuclear pc-scale discs
discussed here.  Recently Seifert \& Scorza (1996) examined in detail
the images of 15 S0 galaxies and found, in a significant fraction of
them, evidence for discs with inner cut-offs, signatures of inner
discs and rings, and kinematic features connected to the double-disc
structure.  Although the resolution of these images is not sufficient
to unambiguously reveal pc-scale nuclear discs, the double disc
structure nevertheless seems to be a common property of early-type
galaxies with kpc-scale discs rather than a mere rarity.  After all,
the double disc structure is only detectable if the galaxy is oriented
sufficiently close to edge-on (Rix \& White 1990).

In order to find constraints on the origin of these complex
disc-structures, it is important to study in detail the photometric
properties of the various disc components once they have been
separated from the bulge via photometric decomposition.  In this paper
we focus on the inner photometric structure of two power-law
early-type galaxies in the Virgo cluster that harbor both an outer
and a nuclear disc: NGC~4342 and NGC~4570.  We separate the nuclear
discs from the inner regions of the bulges in order to study their
photometric properties and the transition region towards the outer
discs.  In particular, we investigate the location of these nuclear
discs in the parameter space defined by the central surface brightness
$\mu_0$ and the scalelength $R_d$ of galactic discs.  The relations
between bulge and disc properties are also investigated in order to
look for constraints on possible formation processes.

Secular evolution, driven by (now dissolved) bars, has been invoked to
explain the formation of nuclear discs, truncated outer discs, and
rings in early-type galaxies (Bagget, Bagget \& Anderson 1996;
Emsellem \etal 1996; van den Bosch \& Emsellem 1998). Small nuclear
bars have been proposed as efficient mechanisms to transport
pre-enriched gas into the centers of galaxies (Friedli \& Martinet
1993; Wada \& Habe 1995).  The fact that S0 galaxies exhibit high
line-strength values at their centers (Fisher, Frank \& Illingworth
1996) and that there is evidence for the existence of embedded bars in
early-type galaxies (e.g., Busarello \etal 1996, Scorza \etal 1998),
supports this secular-evolution picture. However, the available data
is at present insufficient to make any conclusive statements about
the formation of nuclear stellar discs. We will speculate on other
possible formation scenarios in the discussion section of this paper.

This paper is organized as follows. Section~2 discusses the HST data
of NGC ~4342 and NGC~4570.  In Section~3 we use this data to decompose
the galaxies into their bulge and disc components and we discuss the
uniqueness and limitations of the decompositions.  In Section~4 we
present the results of the photometric decomposition and discuss the
properties of the separate components.  In Section~5 we show the
location of the nuclear discs in the $\mu_0$-$R_d$ diagram, and
compare them to discs in spirals, S0s and embedded-discs in
ellipticals. Finally, in Section~6, we discuss our results and
speculate on possible formation scenarios for these multiple disc
structures.

\section{2 The HST images}

\subsection{2.1 The data}

The nuclear stellar discs in NGC~4342 and NGC~4570 were discovered
from Wide Field and Planetary Camera 1 (WFPC1) HST images (van den
Bosch \etal 1994). In addition, the same study revealed a nuclear
stellar disc in the E/S0 NGC~4623. The quality of these data suffers
from the spherical aberration of the HST primary mirror.

In a follow-up program, van den Bosch, Jaffe \& van der Marel (1998,
hereafter BJM98) obtained $U$, $V$ and $I$ band images of NGC~4342 and
NGC~4570 with the Wide Field and Planetary Camera 2 (WFPC2) and the
now refurbished HST. The spatial resolution of these images is set by
the HST Point Spread Function (PSF), which has a FWHM of $\sim 0.1''$,
and by the size of the CCD pixels ($0.0455'' \times 0.0455''$).
Details on the images and their reduction can be found in BJM98.
Because of the better quality of these data we use the WFPC2 $V$-band
images of NGC~4342 and NGC~4570 for our photometric decomposition.
Unfortunately, no WFPC2 images are available for NGC~4623. Because of
the poorer quality of the WFPC1 data we do not include NGC~4623 in our
sample here. 

\subsection{2.2 Deconvolution}

Because of the small sizes of the nuclear discs, and the steep central
luminosity profiles of the galaxies, the PSF can have strong effects
on the outcome of the photometric decomposition. Even though the HST
PSF has improved significantly with the 1993 refurbishment mission,
the PSF has still relatively broad wings with several percent of the
light being scattered more than $1''$ away. It is therefore essential
to deconvolve the images in order to be able to derive the proper
nuclear disc parameters.

BJM98 have presented models of the deconvolved surface brightness of
NGC~4342 and NGC~4570. These models were constructed with the Multi
Gaussian Expansion method developed by Emsellem, Monnet \& Bacon
(1994), and describe the observed (convolved) surface brightness
distribution as a sum of flattened Gaussians. Upon describing the PSF
as a sum of Gaussians as well, one can recover the intrinsic
(deconvolved) surface brightness distribution, also described by a sum
of flattened Gaussians.  In the present work, we use these models to
determine the parameters of the nuclear discs in these two galaxies.
The parameters of the MGE models and the fit of the models to the
observed surface brightness are presented in BJM98.  The MGE
deconvolved surface brightness is in excellent agreement with the
results from Lucy deconvolution (cf. BJM98).

\section{3 Photometric decomposition}

\subsection{3.1 The method}

We apply the method described by Scorza \& Bender (1995) to derive the
parameters of the nuclear discs. This method is particularly well
suited for a photometric decomposition of a system with a disc that
does not strongly dominate the projected surface brightness. The
method is based on the assumption that the bulge component has
perfectly elliptical isophotes.  No assumptions are made regarding the
flattening or the luminosity profile of the bulge. Infinitesimally
thin exponential disc models are subtracted iteratively from the
galaxy frames, until the remaining bulge shows perfectly elliptical
isophotes. The disc model is described by a central surface brightness
$\mu_0$, a scalelength $R_d$, and an inclination angle $i$. In both
galaxies discussed here the nuclear discs could be accurately fitted
with an exponential profile.

Before and after subtraction of the disc models an isophotal analysis
of the galaxy image is carried out using the method described in
Bender \& M\"ollenhoff (1987). In addition to the surface brightness,
ellipticity, and position angle of each isophote, the method
determines the higher-order Fourier coefficients that describe
deviations of the isophote from an elliptical shape. The most
well-known of these is the fourth cosine coefficient $a_4$, which
describes whether the isophote is discy ($a_4 > 0$) or boxy ($a_4 <
0$).  This $a_4$ parameter has proven to be a very good indicator for
the existence of embedded discs (Scorza \& Bender 1995). The best
fitting disc parameters are determined by finding the disc model
which, after subtraction from the galaxy image, yields the smallest
rms values of the $a_4$ Fourier coefficients over the radial interval
where the disc is expected to be visible against the background of the
bulge light.

%
\beginfigure{1}
\centerline{\psfig{figure=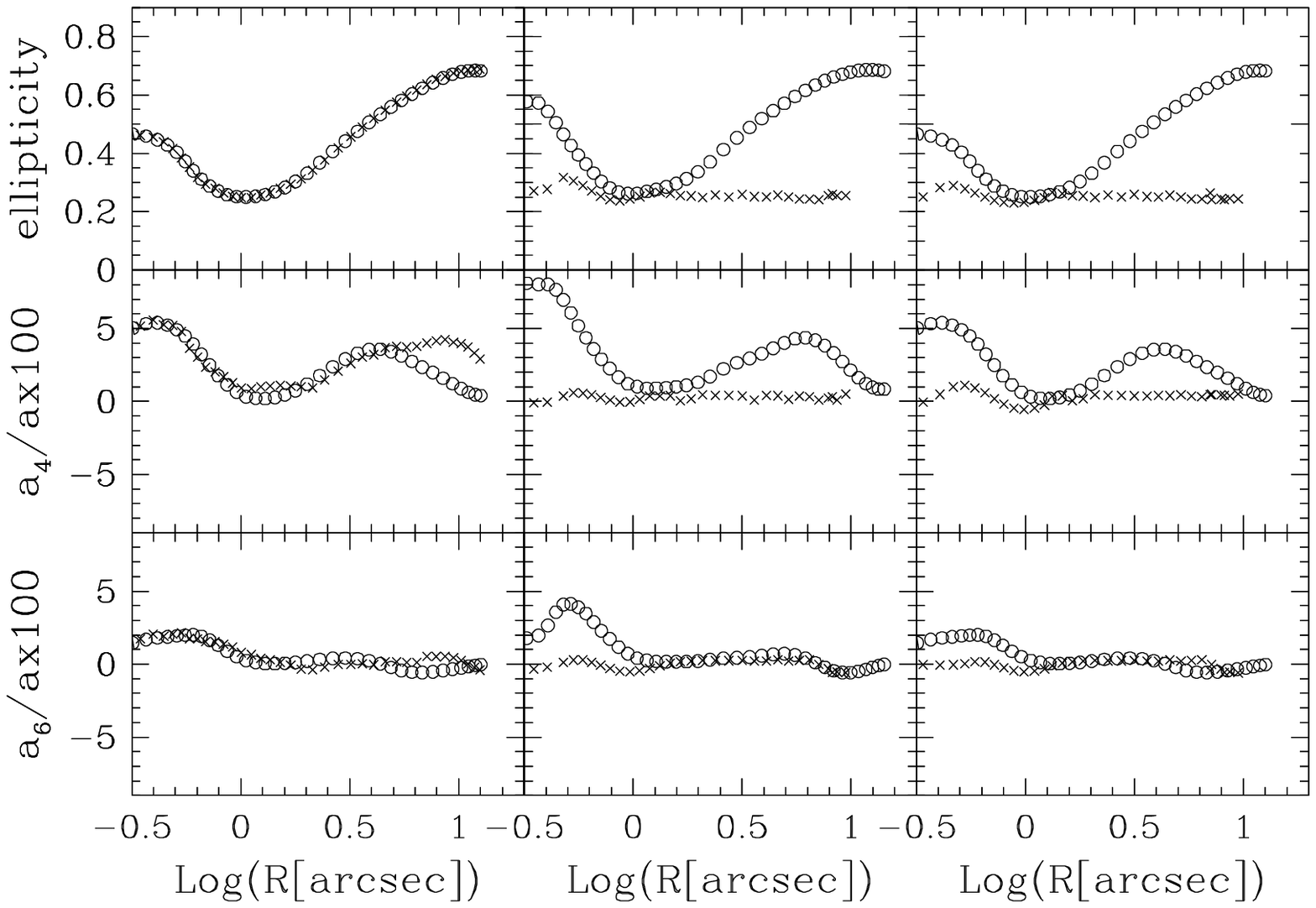,width=\hssize}}\smallskip
\caption{{\bf Figure 1.} Ellipticity, $a_4$ and $a_6$ Fourier 
  coefficient profiles of NGC~4342. The left panels show the profiles
  of the original image (crosses) and the MGE-model (open circles).
  Middle panels: same profiles of the deconvolved MGE-model before
  (open circles) and after (crosses) subtraction of the nuclear and
  outer discs. Right panels: MGE-model prior to deconvolution before
  (open circles) and after (crosses) subtraction of the $\it
  convolved$ nuclear and outer disc models.} \endfigure
%

%
\beginfigure{2}
\centerline{\psfig{figure=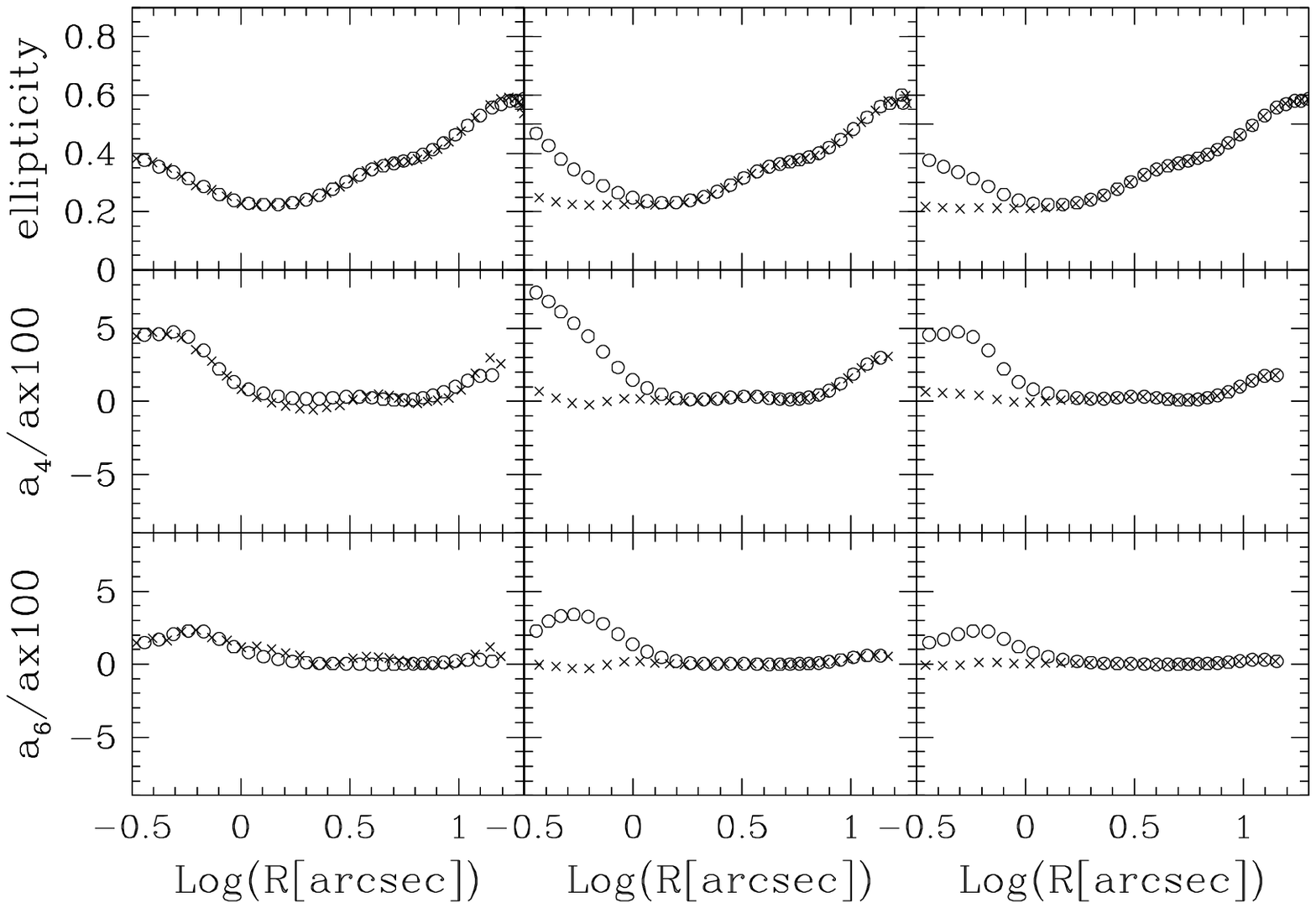,width=\hssize}}\smallskip
\caption{{\bf Figure 2.} Same as Figure~1, but now for NGC~4570.
  Note that for this galaxy, no outer disc parameters could be
  determined (see text).} \endfigure
%

The field-of-view of the HST images used here is unfortunately too
small to allow a proper analysis of the outer disc parameters.  In the
case of NGC~4342, however, the extent of the galaxy expressed in solid
angle on the sky is sufficiently small such that the outer disc of
this galaxy is almost completely registered on the HST image.  Since
the outer disc in NGC~4342 dominates the projected surface brightness
at large radii, it does not predominantly show up in the higher-order
Fourier coefficients, but mainly cause a strong increase in the
ellipticity of the outer isophotes.  For this reason, we cannot
decompose the outer disc from the bulge by using the method described
above.  However, in order to study the transition region and change of
properties between the inner and outer discs, we use another iterative
method to determine the outer disc parameters of NGC~4342.  This
method is described in detail in Scorza \etal (1998),
where it is applied to decompose several disc-dominated early-type
galaxies. The procedure is as follows: after subtraction of an initial
outer disc model, the bulge profile in the intermediate region, where
the bulge dominates the projected surface brightness, is fit by a
$r^{1/4}$-law profile and extrapolated out to radii where the disc
dominates. From this bulge profile, an $r^{1/4}$-bulge model is
constructed and subtracted from the galaxy image; the residual disc is
then used as input disc for a further iteration. In this way, a
reliable decomposition is achieved.

\subsection{3.2 Reliability and uniqueness of the decomposition}

As shown by Rix \& White (1990), several combinations of disc-to-bulge
ratios and inclination angles can produce similar $a_4$ profiles.
Therefore, in order to obtain unique disc solutions from the $a_4$
coefficient, it is necessary to constrain the inclination angle of the
disc. In the method of Scorza \& Bender (1995) this is done via the
$a_6$ Fourier coefficient.  Indeed it has been found that only one
combination of central surface brightness $\mu_0$, scalelength $R_d$,
and inclination $i$ of the disc models leads to simultaneously
vanishing $a_4$ and $a_6$ coefficients and therefore, to a unique disc
solution (see tests in Scorza \& Bender 1990).

It is important to realize that the parameters of the discs determined
from the decomposition methods outlined above are based on the
assumption that the discs are infinitesimally thin.  Realistic discs
have a finite thickness, and taking this thickness into account is
especially important when constructing dynamical models for these
systems. In the Appendix we discuss how to convert the infinitesimally
thin discs to a set of more realistic, thick disc models. We also
discuss the degeneracy between the inclination angle and the thickness
of galactic discs. Thus, whereas the assumption of zero-thickness
yields well defined unique disc parameters, relaxing the assumption
that discs are infinitesimally thin, results in some amount of
non-uniqueness regarding the inclination angle. As long as the galaxy
is seen sufficiently close to edge-on, this non-uniqueness is small
and unimportant.

Recently, in studying the deprojection of axisymmetric bodies, Gerhard
\& Binney (1996) found that the $D/B$ ratios derived from
decomposition procedures are in principal ill-determined from the
photometry alone unless the disc is bright and seen near to edge-on.
This is due to the fact that there are different intrinsic density
distributions that project to the same surface brightness (Rybicki
1986). In the present study we concentrate on close to edge-on,
prominent bright nuclear discs, such that the non-uniqueness of the
deprojected light distribution is only small and the disc-to-bulge
ratio well constrained (cf. Romanowsky \& Kochanek 1997; van den Bosch
1997).
 
\section{4 Results}

\subsection{4.1 The nuclear disc parameters}

The results of the photometric decompositions of NGC~4342 and NGC~4570
are shown in Figures~1 and~2, respectively. Upper panels show the
ellipticity, middle panels show $a_4/a \times 100$, and lower panels
show $a_6/a \times 100$. The left panels of Figures~1 and~2 show the
isophotal parameters of the original HST images (crosses) and the
MGE-models (open circles) prior to deconvolution.  These parameters
agree very well with each other, indicating the accuracy of the fit of
the MGE-models to the data.  A discrepancy in the $a_4$ coefficient at
the outside of the MGE-model of NGC~4342 is visible in Figure~1. This
discrepancy is due to a small isophote twist in the outer region of
this galaxy, not taken into account by the MGE-model (see BJM98), and
does not significantly influence the parameters determined for the
outer disc.  The middle panels of Figures~1 and~2 show the deconvolved
MGE-models before (open circles) and after (crosses) the subtraction
of the best fitting disc models.  As can be seen, after disc
subtraction the central regions of the bulges become elliptical (near
to zero $a_4$ and $a_6$ Fourier coefficients) and of moderate
flattening as the ellipticity drops down to small values.  For
NGC~4342 both the outer and the nuclear discs are subtracted, whereas
for NGC~4570 this is limited to the nuclear disc only.

Once the best-fitting disc model is found, we convolve it with the
same PSF that we used for the deconvolution of the MGE-model, and
subtract this {\it convolved} disc model from the MGE-model prior to
deconvolution.  In the absence of deconvolution artifacts, this should
yield a similar vanishing of the $a_4$ and $a_6$ Fourier coefficients
and a similar ellipticity of the central part of the bulge as the
results from the deconvolved photometry.  This procedure allows us to
cross-check the disc parameters derived.  The results of this analysis
are shown in the right panels of Figures~1 and~2 were we have plotted
the isophotal parameters of the MGE-model prior to deconvolution (same
as in left panels) before (open circles) and after (crosses)
subtraction of the {\it convolved} disc model. As can be seen, the
results in the middle and right panels are in excellent agreement,
indicating that there are no deconvolution artifacts.


\begintable{1}
\caption{{\bf Table~1.} Parameters of nuclear and outer discs}
\halign{
#\hfil&
\quad \hfil#\hfil\quad&
\quad \hfil#\hfil\quad&
\quad \hfil#\quad&
\quad \hfil#\hfil\quad&
\quad \hfil#\hfil\quad&
\hfil#\hfil\quad \cr

NGC & $\mu_0$ & $\mu^{c}_0$ & $R_d$ & $i_0$ & $L_{\rm disc}$ & N/O \cr 
(1) & (2) & (3) & (4) & (5) & (6) & (7) \cr 
\noalign{\medskip\hrule\medskip} 
4342&$13.14$&$15.43$&$7.3$&$83.0$&$8.1 \times 10^6$&N\cr 
4342&$17.39$&$19.10$&$430.0$&$78.0$&$9.7 \times 10^8$&O\cr 
4570&$14.85$&$17.06$&$23.5$&$82.5$&$1.9 \times 10^7$&N\cr 
}
\tabletext{Column~(1) lists the NGC number of the galaxy.  Column~(2)
  gives the projected central surface brightness of the nuclear disc
  in magn arcsec$^{-2}$. The face-on corrected value for this, derived
  using equation (1), is listed in column~(3). Column~(4) gives the
  scalelength in pc, column~(5) the inclination angle (assuming the
  disc is infinitesimally thin), and column~(6) the total luminosity
  of the nuclear discs in $\Lsun$. The last column indicates either a
  nuclear disc `N' or an outer disc `O'.} \endtable


In both galaxies the nuclear discs could be accurately fitted
with an exponential profile.  Table~1 lists the nuclear disc
parameters: the projected central surface brightness of the nuclear
disc $\mu_0$, the central surface brightness of the disc corrected to
face-on $\mu_0^c$ via
$$ \mu_0^c = \mu_0 - 2.5 \log (\cos i), \eqno(1)$$
the scalelengths $R_d$ converted to parsecs via the radial velocities
of the RC3 catalogue (de Vaucouleur \etal 1991) and assuming $H_0 =
100 \kms {\rm Mpc}^{-1}$; the inclination angle $i$ (assuming the disc
is infinitesimally thin), and the total luminosity $L_{\rm disc}$ in
$\Lsun$. Nuclear discs are indicated by `N', outer discs by `O'.

\subsection{4.2 Discussion on individual objects}
 
\noindent
{\bf NGC~4342}

\noindent
This galaxy is classified as E7 and S$0^-$ in the RSA (Sandage \&
Tammann 1981) and RC2 (de Vaucouleur, de Vaucouleur \& Corwin 1976)
catalogues, respectively. The rotation curve of this galaxy reveals
the nuclear disc as a kinematically distinct component (BJM98).  No
strong color gradients are found, especially inside $\sim 2''$, so
that no population difference between bulge and nuclear disc is
apparent (see BJM98). The nucleus of NGC~4342 harbors a massive black
hole (BH) of $\sim 1.4 \times 10^8 \Msun$ (Cretton \& van den Bosch
1998).  The dynamical modeling of these authors also yields an
$I$-band mass-to-light ratio of $13.2$. Both these numbers have been
converted to the distance of NGC~4342 of 7.14 Mpc adopted in this
paper.  Together with the observed central $(V-I)$-color of $\sim
1.3$, and the total $V$-band luminosity of the nuclear disc listed in
Table~1, this yields a total mass of the nuclear disc of $\sim 1.7
\times 10^8\Msun$ (using $(V-I)_{\odot} = 0.81$), comparable to the
mass of the BH.  Figure~3 shows the surface brightness profile
along the major axis of this galaxy (full line), its outer disc
(long-short dashed line), nuclear disc (short dashed line) and bulge
(dashed-dotted line). The subtraction of the outer disc from the
MGE-model of NGC~4342 yields an almost perfectly elliptical bulge (see
$a_4$ profile at large radii in the middle panel of Figure~1).

\bigskip

%
\beginfigure{3}
\centerline{\psfig{figure=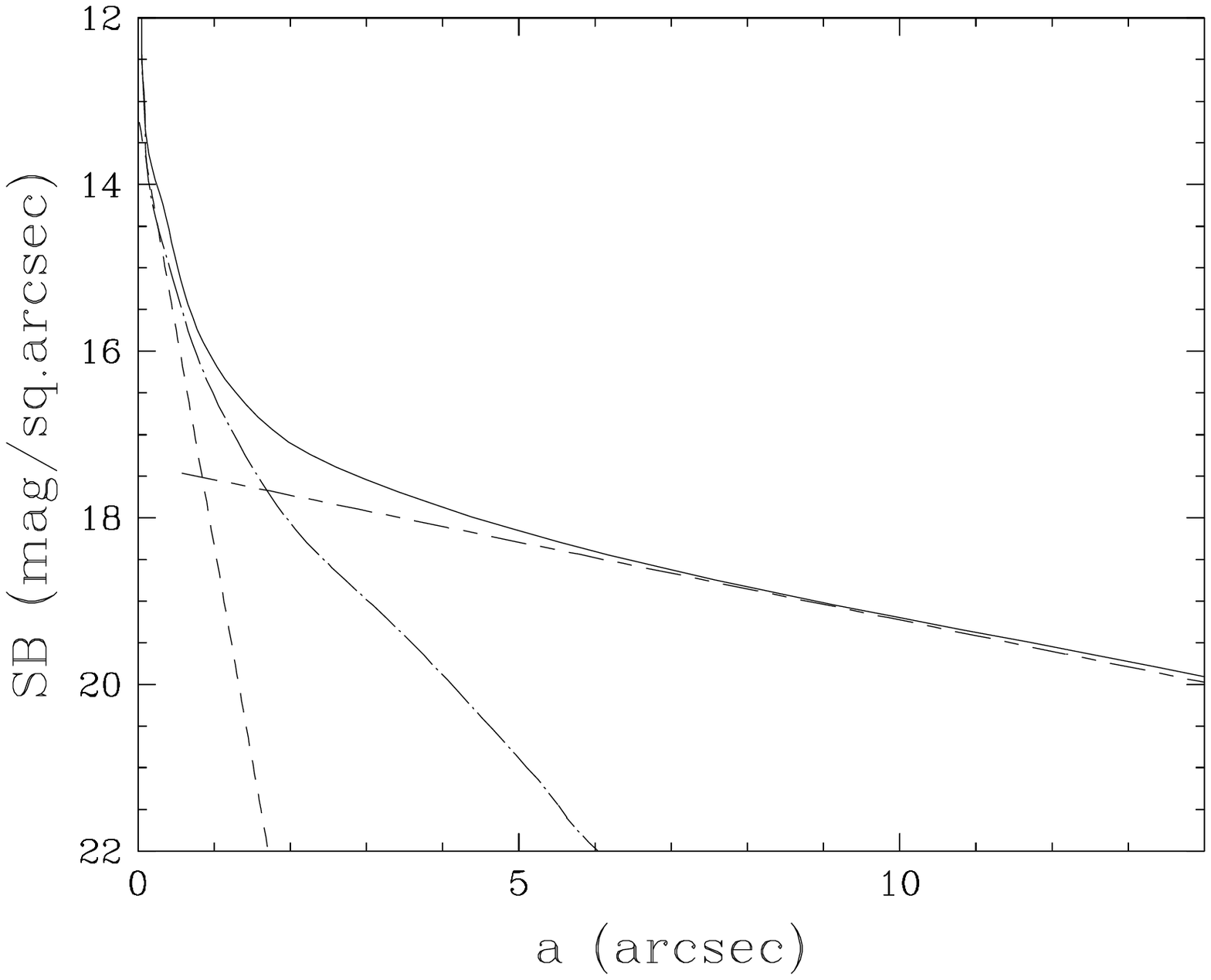,width=\hssize}}\smallskip
\caption{{\bf Figure 3} 
  Surface brightness profile along the major axis of NGC~4342 (full
  line), its nuclear disc (short dashed line), outer disc (short-long
  dashed line) and bulge (dotted-dashed line)} \endfigure
%

\noindent
{\bf NGC~4570}

\noindent
NGC~4570 is classified as S0/E7 and S0 in the RSA and RC2 catalogues
respectively.  Like NGC~4342, its nuclear disc is visible as a
distinct component in the rotation curve. BJM98 report a significant
color gradient in this galaxy and an unusually large $H\beta$
line-strength in the nucleus, suggestive of recent star-formation. Van
den Bosch \& Emsellem (1998) found strong evidence for secular
evolution driven by a nuclear bar in NGC~4570. They report two edge-on
rings between the nuclear and outer discs, whose locations are
consistent with the Inner Lindblad and Ultra Harmonic Resonances of a
tumbling triaxial potential. Because of the small HST field, it was
not possible to investigate the outer disc of this galaxy. Figure~4
shows the surface brightness profile along the major axis of NGC~4570
(full line) and of its nuclear disc (dashed-line).  Due to the small
contribution of the nuclear disc to the central luminosity, the
bulge$+$outer-disc profile in this region (dotted-dashed line) is
hardly distinguishable from the galaxy profile.

%
\beginfigure{4}
\centerline{\psfig{figure=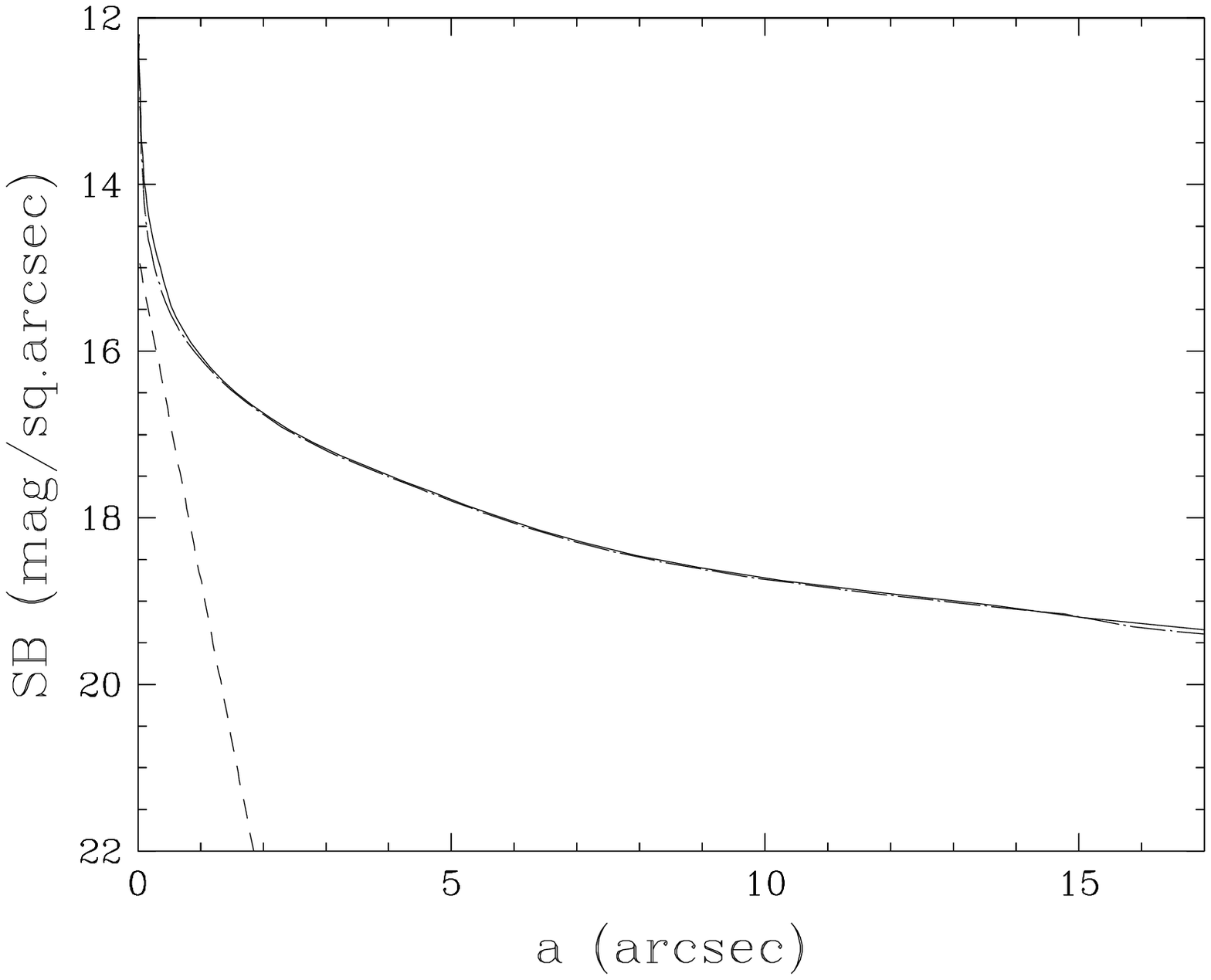,width=\hssize}}\smallskip
\caption{{\bf Figure 4} 
  Surface brightness profile along the major axis of NGC~4570 (full
  line) and its nuclear disc (dashed line). Due to the small
  contribution of the nuclear disc to the central luminosity, the
  bulge$+$outer-disc profile in this region (dotted-dashed line) is
  hardly distinguishable from the galaxy profile} \endfigure
%

\subsection{4.3 Nuclear and outer disc: two discs or one?}

Are the nuclear and outer discs really two separate structures, or are
they merely the manifestation of a single disc that has a
double-exponential surface brightness profile?  Since NGC~4342 is the
only case for which we could determine the disc parameters of both the
outer and the nuclear disc, we focus here on this galaxy only.  As can
be seen from Table~1, the outer and nuclear discs in NGC~4342 seem to
indicate a different inclination angle: the nuclear disc yields
$i=83\deg$, whereas from the outer disc we derive $i = 78\deg$.  We
recall that the applied decomposition technique allows a determination
of the inclination angle with a precision of $\sim 2\deg$ (see Scorza
\& Bender 1990). Since we consider it unlikely that both discs are
seen under different inclination angles (as this would imply an
unstable situation), this indicates that the two discs have different
thicknesses (see the Appendix). If we assume that NGC~4342 is seen
edge-on, we find that the outer disc is almost a factor three thicker
than the nuclear disc.  This, strongly suggests that the nuclear and
outer discs are actually two distinct discs rather than a single
double-exponential disc.  The difference in thicknesses suggests
different amounts of energy dissipation during their respective
formations.

As discussed in Section~1, a general feature of multi-disc systems
seems to be the presence of an inner cut-off in the outer disc. The
presence, or absence of an inner cut-off is of particular interest,
since it may provide clues to the formation of the double-disc
structure (see Section~6). As can be seen from Figure~3 it is
unfortunately impossible to discriminate between the presence or
absence of an inner cut-off based on our photometry: the contribution
from the outer disc to the projected surface brightness is completely
negligible in the inner 1 to 2 arcsec, where the bulge and nuclear
disc dominate the light. We have thus determined the outer disc
parameters of NGC~4342 assuming no inner cut-off, but keep in mind
that our current data can not distinguish whether the outer disc of
NGC~4342 harbors an inner cut-off radius, or whether it continues all
the way to the center.


\beginfigure*{5}
\line{\psfig{figure=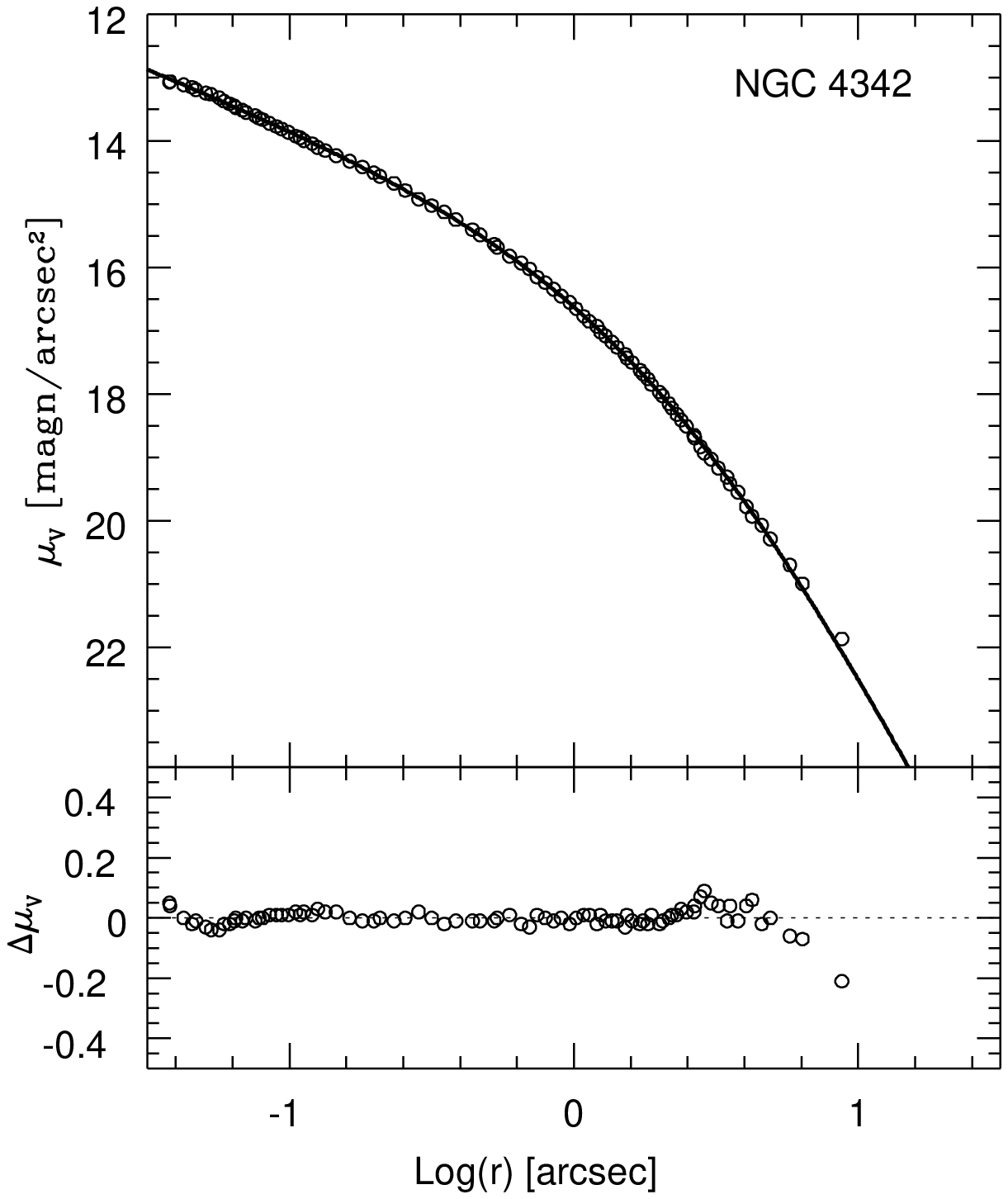,width=\hssize}\hfill
\psfig{figure=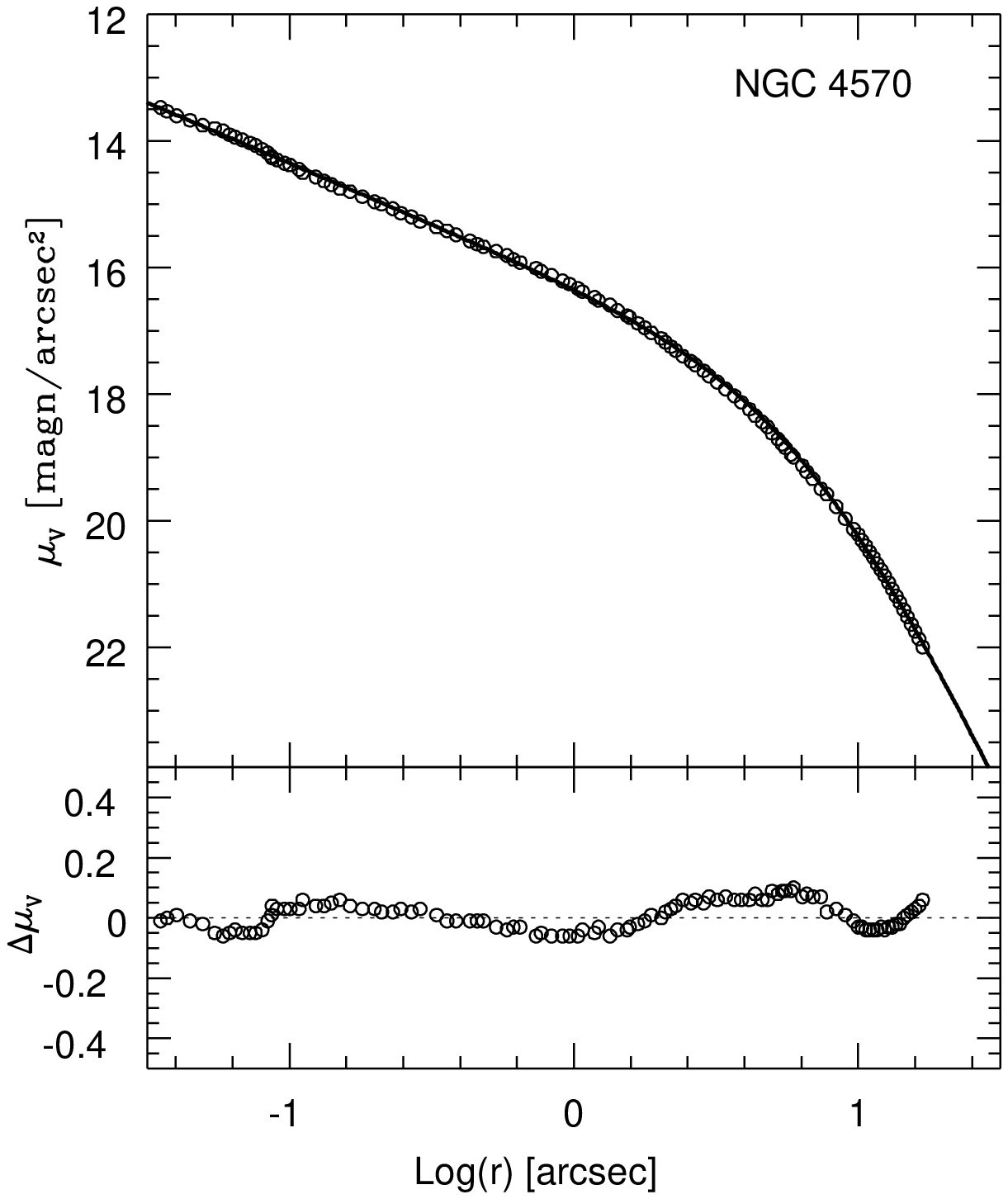,width=\hssize}}\smallskip
\caption{{\bf Figure 5} The Nuker-law fits (thick solid curves) to the
  minor axis bulge profiles (open circles) of NGC~4342 and NGC~4570.
  The nuclear discs have been subtracted from these profiles (see
  text).  The lower panels show the residuals. The Nuker-law fits the
  luminosity profiles remarkably well, yielding residuals no larger
  than $0.1$ magn throughout.} \endfigure


\subsection{4.4 The bulge parameters}

Photometry with the HST has revealed that the central luminosity
profiles of early-type galaxies are cusped, and that the galaxies can
be classified in two classes according to the cusp steepness (see
discussion in Section~1). Here we investigate the luminosity profiles
of the bulges in the systems with a nuclear disc. Lauer \etal (1995)
introduced the so-called ``Nuker''-law profiles, which provide a good
fit to the observed central luminosity profiles of early type
galaxies.  The Nuker law is given by
$$ I(r) = I_b \, 2^{{\beta - \gamma \over \alpha}} \biggl({r\over r_b}
\biggr)^{-\gamma} \Biggl[ 1 + \biggl({r\over r_b}\biggr)^{\alpha}
\Biggr]^{{\gamma - \beta \over \alpha}}.\eqno(2)$$
Note that for $r \ll r_b$, $I \propto r^{-\gamma}$, while for $r \gg r_b$,
$I \propto r^{-\beta}$. The parameter $\alpha$ controls the sharpness
of the transition from cusp to outer power-law at the break-radius $r_b$.

After subtraction of the nuclear discs from the images, we determine
the luminosity profiles of the bulges along the minor axes, by fitting
isophotes to the disc-subtracted images.  This axis is least affected
by any uncertainty in either the nuclear or the outer disc parameters.
We fit equation (2) to the bulge luminosity profiles using the
Levenberg-Marquardt method to determine the best fitting parameters
$(I_b,r_b,\alpha,\beta,\gamma)$.


\begintable{2}
\caption{{\bf Table~2.} Bulge parameters}
\halign{
#\hfil&
\quad \hfil#\hfil\quad&
\quad \hfil#\hfil\quad&
\quad \hfil#\hfil\quad&
\quad \hfil#\hfil\quad&
\quad \hfil#\hfil\quad&
\hfil#\hfil\quad \cr

NGC & $I_b$ & $r_b$ & $r_b/R_d$ & $\alpha$ & $\beta$ & $\gamma$ \cr 
(1) & (2) & (3) & (4) & (5) & (6) & (7) \cr 
\noalign{\medskip\hrule\medskip} 
4342 & $18.28$ & $2.28$ &  $8.14$ & $1.03$ & $3.56$ & $0.73$ \cr 
4570 & $19.41$ & $7.34$ & $26.21$ & $1.57$ & $3.88$ & $0.77$ \cr }

\tabletext{Parameters of the Nuker-law (eq. [2]) that best fit the
  minor axis bulge luminosity profiles of NGC ~4342 and NGC~4570.
  Column~(1) lists the NGC number of the galaxy. Column~(2) gives the
  surface brightness $I_b$ in magn arcsec$^{-2}$ at the break radius
  $r_b$, which is listed, in arcsec, in column~(3). Column~(4) gives
  the ratio of the break radius of the bulge over the scalelength of
  the nuclear disc $R_d$. Columns (5), (6), and (7) list the
  parameters $\alpha$, $\beta$, and $\gamma$.} \endtable


Figure~5 shows the minor axis luminosity profiles of the bulges (open
circles), along with our best fitting Nuker-laws (solid lines). The
lower panels show the residuals.  As can be seen, the fits are very
good, yielding residuals less than 0.1 magnitudes throughout. The best
fitting parameters are listed in Table~2.  Both bulges have very steep
cusps ($\gamma \sim 0.7$ - $0.8$) and harbor a clear break (i.e.,
large difference between $\beta$ and $\gamma$).

There has been some discussion in recent literature as to the origin
of the light that produces the steep cusp in these galaxies.  It was
suggested by Jaffe \etal (1994) to be due to the nuclear disc seen
close to edge-on. Faber \etal (1997) investigated a much larger sample
of early-type galaxies imaged by HST, and concluded that it is the
spheroidal component which has an intrinsic steep cusp.  The analysis
performed here clearly reveals that indeed the bulges in NGC~4342 and
NGC~4570 are steeply cusped, and that the contribution to the central
light by the nuclear disc is only small compared to that of the
bulge's cusp.

\section{5 The $\mu_0$--$R_d$ diagram}

When Freeman (1970) studied the photometry of 36 spiral galaxies, he
found the amazing result that the central surface brightness of 28
galaxies of this sample fell within the range of $\mu_0(B) = 21.65 \pm
0.3$ magn.  arcsec$^{-2}$. Ever since, there has been a considerable
debate on the validity of this so-called Freeman law. Some showed that
the effect might be real (e.g., van der Kruit 1987), while others
argued that it is due to selection effects (e.g., Disney 1976) and/or
extinction (Valentijn 1990).  The discovery of low surface brightness
(LSB) galaxies in a number of surveys (e.g., Schombert \& Bothun 1988;
Impey, Bothun \& Malin 1988; Schombert et al. 1992; McGaugh \& Bothun
1994; de Blok, van der Hulst \& Bothun 1995) as well as recent
near-infrared photometry studies of large samples of `normal' or
so-called high surface brightness (HSB) galaxies (e.g., de Jong \& van
der Kruit 1994) have clearly revealed a large range in central surface
brightness amongst disc galaxies, contrary to Freeman's law.  An
extreme example is Malin I (Impey \etal 1988; Impey \& Bothun 1989),
which is the largest (LSB) spiral galaxy known to date and has a central
surface brightness of $\mu_0(B) = 26.5$ magn. arcsec$^{-2}$, almost 5
magnitudes fainter than Freeman's value.  When plotting the central
surface brightnesses $\mu_0$, versus the scalelengths $R_d$ of the
discs, the LSB extend the region occupied by HSB discs but do not fill
in the entire range between normal spirals and Malin 1 (Sprayberry
\etal 1995).  Thus, among spirals galaxies there seems to be a clear
upper limit to the surface brightness of the discs, but no clear limit
at the faint end of the $\mu_0$ distribution.

If the parameters of the embedded discs in discy-ellipticals are
included in the $\mu_0(V)$--$R_d$ parameter space (Scorza \& Bender
1995), it is found that these discs extend the region occupied by
spiral and S0 discs towards higher $\mu_0$ and shorter scalelengths
$R_d$. The nuclear discs studied here reach even higher central
surface brightnesses and have much smaller scalelengths.  In Figure~6
we plot the central surface brightness (in $V$-band) as a function of
the logarithm of the scalelength of a vast variety of discs, ranging
from the extraordinary large disc of Malin I (solid triangle, data
taken from Bothun \etal 1987), to the nuclear discs studied in this
paper (asterisks). We include the parameters of the nuclear disc in
NGC~3115 (open triangle) as derived by Kormendy \etal (1996a) with the
same method as used here.  In addition, we plot the parameters of
discs in S0s and embedded discs in discy ellipticals (solid circles,
data taken from Scorza \& Bender 1995, and Scorza \etal 1998), and of
a combined sample of HSB and LSB spiral discs (open circles, data
taken from Kent 1985; de Jong 1996; Sprayberry \etal 1995; de Blok,
van der Hulst \& Bothun 1995; McGaugh \& Bothun 1994). For comparison,
we have also plotted the Freeman value.  Figure~6 presents the results
for discs that span almost 4 orders of magnitude in both scalelength
and central surface brightness.  Clearly, the nuclear discs are as
extreme in their photometric properties as Malin I, when compared to
typical `normal' HSB spiral discs that obey Freeman's law.

It is important to note here that the main parameter that changes
along the sequence of discs depicted in the diagram of Figure~6 is the
disc-to-bulge ratio $D/B$. Spirals and S0s typically have $D/B \gta
1$, discy ellipticals have $D/B \sim 0.1$, and the nuclear disc
systems studied here have $D/B \sim 0.01$.  The importance
of this parameter with regard to disc stability and the position of
the different discs in the $\mu_0(V)$--$R_d$ diagram is discussed in
van den Bosch (1998).
 

\beginfigure{6}
\centerline{\psfig{figure=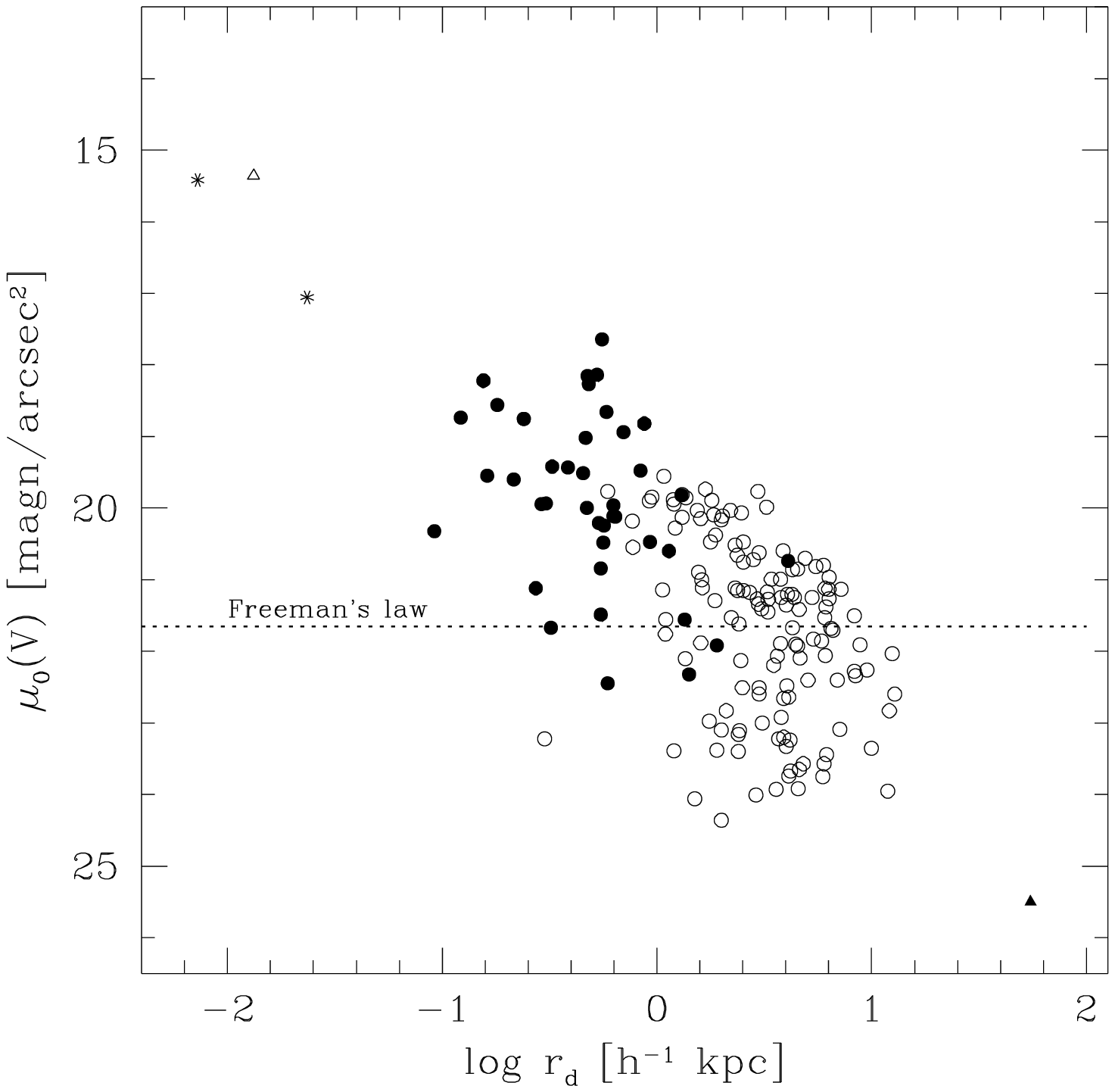,width=\hssize}}\smallskip
\caption{{\bf Figure 6} 
  $\mu_0$ and $R_d$ diagram of a vast variety of discs, including
  Malin I (solid triangle, taken from Bothun \etal 1987), discs in
  spirals (open circles, data taken from Kent 1985; McGaugh \& Bothun
  1994; de Jong 1996; Sprayberry \etal 1995; de Blok, van der Hulst \&
  Bothun 1995), discs in S0s and discy ellipticals (filled circles,
  data taken from Scorza \& Bender 1995 and Scorza \etal 1998), the
  nuclear disc in NGC~3115 (open triangle, taken from Kormendy \etal
  1996a) and the nuclear discs examined here (asterisks). 
  For comparison the Freeman value is plotted as a dotted
  line.} \endfigure
%

\section{6 Conclusions}

We have studied the photometric structure of two power-law early-type
galaxies with nuclear stellar discs. Special attention has been paid
to the nuclear disc components, which were separated from the central
part of the bulges by means of a photometric decomposition method.  We
find a continuity of photometric properties in the parameter space
defined by the central surface brightness $\mu_0$ and the scalelength
$R_d$ of discs in LSB, HSB spirals, S0s and embedded discs in
ellipticals, in the sense that the nuclear discs extend the observed
disc properties even further towards smaller scalelengths and brighter
central surface brightnesses. The $\mu_0$--$R_d$ diagram includes now
discs that cover four orders of magnitude in both scalelength and
central surface brightness. The nuclear discs studied here are the
smallest and brightest stellar discs known, and as such, they are as
extreme in their properties as Malin I, when compared to typical
galactic discs that obey Freeman's law.  The small scalelengths of
these discs ($R_d < 25$ pc) and their relatively large masses ($\sim
10^8 \Msun$) suggest strong dissipational processes in the central
regions of their host galaxies.

Nuker-law fits to the surface brightness profiles of the bulge
components of NGC~4342 and NGC~4570 along the minor axis reveal that
the bulges have very steep cusps. We thus confirm the conclusion of
Faber \etal (1997) that the central surface brightness in these
galaxies is dominated by that of the bulges and not by the nuclear
discs. The latter only contribute significantly to the projected
surface brightness in a limited radial interval between $\sim 0.3''$
and $\sim 1.0''$.

In recent years it has become apparent that many disc galaxies have
double-disc structures. This discovery has prompted the
question as to the origin of this multi-component disc structure.  One
possible formation mechanism is related to bars. In particular, it has
been argued that the inner-truncated discs could be linked with the
presence of a tumbling bar potential (Baggett, Baggett \& Anderson
1996). In fact, in one of the galaxies studied here, NGC~4570, there
is strong evidence that the multi-disc structure has indeed been
shaped by secular evolution induced by a small bar (van den Bosch \&
Emsellem 1998). A similar conclusion was reached by Emsellem \etal
(1996) for the case of the Sombrero galaxy, which is the prototypical
galaxy with a double disc structure. Thus, secular evolution driven by
a nuclear bar could be the process responsible for the formation of
nuclear disc components in power-law early-type galaxies. One of the
promising features of this scenario is that it provides a natural
explanation for the double disc structure observed in most, if not
all, cases where nuclear stellar discs have been discovered: the outer
disc became bar unstable, and the nuclear disc formed out of
(pre-enriched) gas transported inwards by this bar structure. In this
scenario, one expects an inner cut-off of the outer discs at radii
similar to the semi-major axis of the nuclear bar.  The significant
difference between the inclination angle inferred from the different
disc components, suggests that the nuclear and outer discs are two
distinct discs, rather than merely the manifestation of a single disc
with a double-exponential surface brightness profile. Whether or not
the outer discs harbor an inner cut-off can unfortunately not be
addressed by the current data.

Although the bar-hypothesis outlined above seems promising, we can
currently not rule out alternative formation scenarios for the nuclear
discs. One of these alternatives may be related to the formation of
central BHs. A significant fraction of the galaxies with nuclear
stellar discs have been shown to harbor massive BHs (e.g., NGC~3115,
Kormendy \etal 1996a; NGC~4342, Cretton \& van den Bosch 1998;
NGC~4594, Kormendy \etal 1996b).  It has to be noted however that this
may be an observational bias since the presence of a nuclear disc
helps in the detection of a BH (see e.g., van den Bosch \& de Zeeuw
1996; Ford \etal 1997). The mass of the nuclear disc in NGC~4342 is
similar to that of its BH (see Section~4.2).  The same holds for
NGC~4594, for which we have estimated a nuclear disc mass of $1.9
\times 10^9 \Msun$, based on the disc parameters of Emsellem \etal
(1996) and $(M/L)_V=4$ (Kormendy 1988). The BH mass in this galaxy is
$1.0 \times 10^9 \Msun$ (Kormendy \etal 1996b).  For NGC~3115, the
nuclear disc mass of $2.3 \times 10^8 \Msun$ is approximately a factor
four smaller than the BH mass of $1.0 \times 10^9 \Msun$ determined by
Kormendy \etal (1996a). Note that again we have converted all masses
to a distance scale with $H_0 = 100 \kms {\rm Mpc}^{-1}$.  Loeb \&
Rasio (1994) simulated the collapse of primordial gas clouds at a
scale $\lta 1$ kpc.  Although the major fraction of the gas fragments
to form a spheroidal component, typically $\sim 5$ percent of the
initial mass settles to form a smooth nuclear gaseous disc.  This disc
requires a seed BH of mass $\gta 10^6 \Msun$ in order to be stable.
If this seed BH is indeed present (see Loeb \& Rasio (1994) and
references therein for formation scenarios for such seed BH), it can
subsequently grow by steady accretion from this nuclear disc to reach
a typical quasar BH mass $\sim 10^8 \Msun$. Although this scenario
gives account primarily of the formation of quasar BHs, it could
provide a possible link between the nuclear stellar discs that we
observe today and BHs.  The fact that the observed BHs and nuclear
discs have masses that are remarkably similar at least implies that
the nuclear discs in these galaxies could have been massive enough in
the past to be an accretion source for the BHs. Nuclear discs could
thus be the remnants of the primordial gas discs that fed the (seed)
BH.

The above mentioned formation scenarios constitute an important frame
of work for future observations. At the present, the small number of
objects with high resolution photometry and spectroscopy does not
allow to draw definitive conclusions. 

\section{Acknowledgments}

\tx We are grateful to Eric Emsellem for his help with making the
MGE-fits to our images, and to the referee, Dr. D. Carter, for his
helpful comments.  FCvdB is thankful for the hospitality of the
Landessternwarte during his stay, when part of this work was done.  CS
was supported by the DFG (Sonderforschungsbereich 328).  FCvdB was
supported by a Hubble Fellowship, {\#}HF-01102.11-97A, awarded by
STScI.

\section{Appendix: Conversion to thick disc models}

The decomposition method described in Section 3.1 yields the disc
parameters under the assumption that the discs are infinitesimally
thin. This is however not very physical and can result in some
mathematical difficulties when constructing dynamical disc models.  In
this Appendix we show how to convert the infinitesimally thin
exponential discs to a set of thick discs that have an identical
surface brightness distribution as the thin exponentials, and in the
mean time are differentiable in all directions, and therefore ideally
suited for dynamical modeling.

The luminosity density of an infinitesimally thin exponential disc in
the equatorial plane ($z=0$) is
$$ \Sigma^{*}(R) = \Sigma^{*}_0 \exp(-R/R^{*}_d),\eqno(A.1)$$
When projected on the sky under an inclination angle $i_0$, one
obtains the surface brightness distribution
$$S^{*}(x,y)={\Sigma^{*}_0 \over \cos i_0} \exp(-m^{*}/R^{*}_d),\eqno(A.2)$$
where $m^{*} = \sqrt{x^2 + y^2/\cos^2 i_0}$, and $(x,y)$ are coordinates in
the plane of the sky.


\beginfigure{7}
\centerline{\psfig{figure=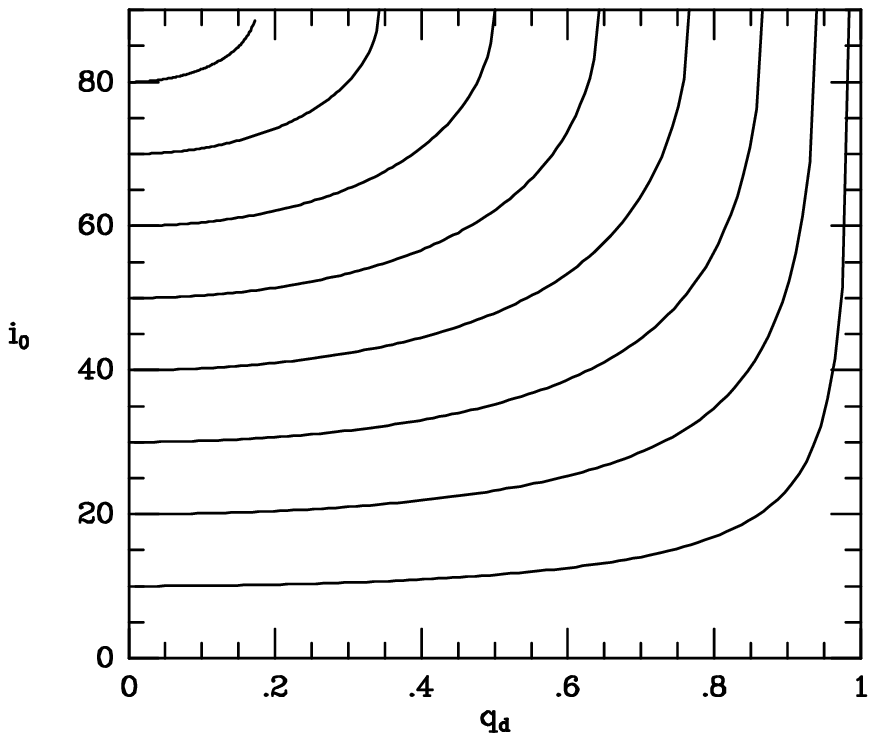,width=\hssize}}\smallskip
\caption{{\bf Figure A.1} The degeneracy between disc flattening $q_d$
and inclination angle $i$. All exponential spheroids (eq. [A.3])
with parameters ($q_d$,$i$) on one of these lines project to exactly 
identical surface brightness, and can therefore not be distinguished 
from photometry alone.} 
\endfigure

        
Van den Bosch \& de Zeeuw (1996) have discussed the so-called
exponential spheroid discs. These are {\it not} infinitesimally thin,
and also project to an exponential surface brightness distribution.
The luminosity density of the exponential spheroid is
$$\nu(R,z) = {\Sigma_0 \over \pi R_d} K_0\bigl({m \over R_d}\big),\eqno(A.3)$$
where $m = \sqrt{R^2 + z^2/q_d^2}$ and $K_0$ is the modified Bessel
function.  In the following we refer to $q_d$ as the thickness of the
disc.  The total luminosity of the exponential spheroid disc is
$$L_{\rm disc} = 2 \pi q_d \Sigma_0 R_d^2.\eqno(A.4)$$
A simple, single quadrature equation for the potential of the exponential
spheroid is given in van den Bosch \& de Zeeuw (1996).
For an inclination angle $i$ the projected surface brightness is given
by
$$S_{i}(x,y) = {q_d \over q'_d} \Sigma_0 \exp(-m'/R_d).\eqno(A.5)$$
Here $m' = \sqrt{x^2 + y^2/q'^2_d}$ and $q'_d$ is the projected
flattening of the disc, which is related to the intrinsic flattening
$q_d$ and the inclination angle $i$ through
$$q'^2_d = \cos^2 i + q^2_d \sin^2 i.\eqno(A.6)$$
It is straightforward to show that when $R_d = R^{*}_d$, 
$$i = {\rm arcsin}\Biggl({\sin i_0 \over \sqrt{1 - q^2_d}}\Biggr),\eqno(A.7)$$
and
$$ \Sigma_0 = {\Sigma^{*}_0 \over q_d},\eqno(A.8)$$
the full 2D surface brightness of the exponential spheroid disc is
exactly equal to that of the infinitesimally exponential disc.
Therefore, one can use the assumption of infinite thinness for the
disc, derive its parameters ($\Sigma^{*}_0$,$R^{*}_d$,$i_0$) from the
disc/bulge decomposition, and subsequently rescale these parameters to
a set of thick discs with $i_0 \leq i \leq 90\deg$ and $0 \leq q_d
\leq \cos i_0$.

This also means that there is a strong degeneracy between inclination
angle $i$ and disc flattening $q_d$. This degeneracy is plotted in
Figure~A.1, for different values of $i_0$ (i.e., the inclination angle
corresponding to the infinitesimally thin disc). All discs with
parameters ($q_d$,$i$) that fall on a curve in Figure~A.1 project to
{\it exactly} identical surface brightness. Therefore, there is no
possibility to uniquely determine both $q_d$ and $i_0$ from the
surface photometry alone.

\bye